\documentclass[sigconf,nonacm]{acmart}
\renewcommand\footnotetextcopyrightpermission[1]{}
\AtBeginDocument{%
  }

\usepackage{tikz}

\usepackage{algorithmic}
\usepackage{graphicx}
\usepackage{textcomp}
\usepackage{bmpsize}
\usepackage{xcolor}
\definecolor{darkgreen}{rgb}{0.0, 0.4, 0.0} 
\definecolor{darkred}{rgb}{0.6, 0.0, 0.0}
\usepackage{lipsum}
\usepackage{etoolbox}
\usepackage{svg}
\usepackage{multirow}
\usepackage{algorithm}
\usepackage{array}
\usepackage{subfigure}
\usepackage{stfloats}
\usepackage{fancybox} 
\usepackage{url}
\usepackage{hyperref}
\usepackage{xurl}
\usepackage{tikz}
\usepackage{fmtcount} % for \ordinalnum
\usepackage{multirow}
\usepackage{multicol}
\usepackage{booktabs}
\usepackage{tabularx}
\usepackage{makecell}
\newcolumntype{C}[1]{>{\centering\arraybackslash}m{#1}}
\newcolumntype{L}{>{\centering\arraybackslash}m{3cm}}

\usepackage{color}

\newcommand*\circled[1]{\tikz[baseline=(char.base)]{\node[shape=circle,draw,inner sep=0.5pt, font=\tiny] (char) {#1};}}

\usepackage{tikz}
\newcommand{\circlednumblack}[1]{
    \tikz[baseline=(char.base)]{
        \node[shape=circle,draw=black,fill=white,text=black,inner sep=1pt] (char) {#1};}
}

\newcommand{\circlednumgray}[1]{
    \tikz[baseline=(char.base)]{
        \node[shape=circle,draw=black,fill=gray,text=black,inner sep=1pt] (char) {#1};}
}
\begin{document}

\title{ThermoCAPTCHA: Privacy-Preserving Human Verification with Farm-Resistant Traceable Tokens}
%------------------------------------------------------------------------------------
%%
% --- Authors (replace the sample author section with this) ---

\author{Shovon Paul}
\affiliation{%
  \institution{School of Computing \& Informatics, University of Louisiana at Lafayette}
  \city{Lafayette}
  \state{Louisiana}
  \country{USA}
}
\email{shovon.paul1@louisiana.edu}

\author{Md Imran Hossen}
\affiliation{%
  \institution{University of Tennessee at Chattanooga}
  \city{Chattanooga}
  \state{Tennessee}
  \country{USA}
}
\email{MdImran-Hossen@utc.edu}

\author{Xiali Hei}
\affiliation{%
  \institution{School of Computing \& Informatics, University of Louisiana at Lafayette}
  \city{Lafayette}
  \state{Louisiana}
  \country{USA}
}
\email{xiali.hei@louisiana.edu}


\begin{abstract}
%------------------------------------------------------------------------------------

CAPTCHAs remain a critical defense against automated abuse, yet modern systems suffer from well-known limitations in usability, accessibility, and resistance to increasingly capable bots and low-cost CAPTCHA farms. Behavioral and puzzle-based mechanisms often impose cognitive burdens, collect extensive interaction data, or permit outsourcing to human solvers. In this paper, we present \textsc{ThermoCAPTCHA}, a novel privacy-preserving human verification system that uses real-time thermal imaging to detect live human presence without requiring users to solve challenges. A lightweight YOLOv4-tiny model identifies human heat signatures from a single thermal capture, while cryptographically bound traceable tokens prevent forwarding attacks by CAPTCHA farm workers. Our prototype achieves 96.70\% detection accuracy with a 73.60\,ms verification latency on a low-powered server. Comprehensive security evaluation—including MITM manipulation, spoofing attempts, adversarial perturbations, and misuse scenarios—shows that ThermoCAPTCHA withstands threats that commonly defeat behavioral CAPTCHAs. A user study with 50 participants, including visually challenged users, demonstrates improved accuracy, faster completion times, and higher perceived usability compared to reCAPTCHA~v2.

\end{abstract}




\maketitle
% 

%------------------------------------------------------------------------------------
\section{Introduction}
%------------------------------------------------------------------------------------
CAPTCHAs serve as a foundational defense against automated abuse on the modern web. From preventing fraudulent account registrations to securing e-commerce checkout flows, CAPTCHAs act as a barrier between legitimate users and increasingly capable automated bots. The original principle behind CAPTCHA design was straightforward: pose a task that is easy for humans but difficult for machines~\cite{von2004telling}. Yet the rapid evolution of machine learning, large-scale CAPTCHA-solving services, and privacy concerns has fundamentally challenged this premise, exposing long-standing weaknesses in widely deployed systems.

Traditional text- and image-based CAPTCHAs exemplify this erosion. Distorted alphanumeric text, once a strong defense, is now easily defeated by modern deep learning techniques~\cite{wang2023experimental}. Image-selection CAPTCHAs---such as identifying traffic lights or crosswalks---became industry standards, but they too are increasingly vulnerable to automated solvers~\cite{aadhirai2012image,hernandez2017oracle}. These challenges also impose notable usability burdens: visually challenged users often cannot complete image-based puzzles, and cultural differences may hinder correct interpretation of visual content~\cite{hernandez2010pitfalls}. Performance degradation and increased page latency further undermine user experience~\cite{ReCAPAdi4:online}. 

Behavioral CAPTCHAs such as reCAPTCHA v2 and v3 attempt to improve usability by replacing explicit puzzles with passive behavioral analysis. These systems infer user legitimacy from mouse movements, timing signals, device fingerprints, or browser patterns~\cite{GoogleOn0:online}. While they reduce user friction in many cases, they introduce significant privacy concerns~\cite{reCAPTCHA35Online,reCAPTCHA76Online}. Behavioral profiling enables cross-site tracking, raises regulatory implications, and may inadvertently discriminate against users with motor impairments or atypical interaction patterns~\cite{guerar2021gotta,hossen2020object}. Moreover, machine learning models have begun mimicking human-like interaction traces, narrowing the security gap originally promised by behavioral approaches.

An equally persistent challenge is the rise of CAPTCHA farms--organized human-solver operations offering bulk CAPTCHA solving for a few dollars per thousand tasks~\cite{narayanaswamy2022captchas,kweuCap8Online,CaptchaS37Online,motoyama2010re,nguyen2024c}. Farm workers solve CAPTCHA challenges remotely, return the solution token to the attacker, and enable automated misuse of online services. Because existing CAPTCHA designs do not cryptographically bind verification tokens to a specific client environment, such tokens can be freely forwarded or reused. This weakness severely limits the effectiveness of behavioral and puzzle-based CAPTCHAs and remains a critical gap in the design of modern web defenses.

These limitations collectively raise a central research question:  
\textit{Can we design a CAPTCHA mechanism that avoids cognitive burden, minimizes privacy risk, resists token forwarding by CAPTCHA farms, and remains accessible to users with visual or auditory impairments?}

\bigskip
\noindent\textbf{This work presents \textsc{ThermoCAPTCHA}, a new direction for CAPTCHA design.}  
ThermoCAPTCHA eliminates puzzle solving and behavioral profiling by authenticating users through a single real-time thermal image capture. Thermal images encode only coarse heat distributions, rendering them non-identifying and inherently privacy-preserving~\cite{lin2021thermal, saad2023hotfoot}. This property also mitigates traditional spoofing attacks that exploit printed photos or RGB replay attempts. A lightweight YOLOv4-tiny model analyzes a single thermal frame to determine whether it contains a live human heat signature, enabling a one-click CAPTCHA experience that requires minimal cognitive effort and supports users with vision or hearing impairments.

Beyond user verification, ThermoCAPTCHA integrates a cryptographically enforced token architecture designed to resist CAPTCHA farm outsourcing. Each verification request includes a signed hash of thermal metadata containing a nonce and timestamp, ensuring integrity and freshness. The CAPTCHA server issues a response token encrypted under both a server secret key and the target website’s shared key. The token embeds a session identifier and device fingerprint, allowing the server to reject any token forwarded from an external environment or reused outside its validity window. This binding property—absent from existing CAPTCHA systems—prevents attackers from outsourcing challenges to remote solvers, restoring a key security guarantee.

We implement a complete end-to-end prototype to evaluate ThermoCAPTCHA across effectiveness, security, and usability dimensions. Empirically, ThermoCAPTCHA achieves 96.70\% human-detection accuracy with an average verification latency of 73.60\,ms on a low-powered server, demonstrating practicality for real-time website integration. Our robustness evaluation spans diverse conditions including varying capture distances, camera angles, lighting environments, and cluttered thermal backgrounds. Security analyses show that ThermoCAPTCHA withstands man-in-the-middle manipulation, replay attempts, tampering with digital signatures, thermal spoofing using heated objects or mannequins, and adversarial perturbation attacks.

To assess real-world usability, we conducted a study with 50 participants, including 20 visually challenged users. ThermoCAPTCHA outperformed reCAPTCHA~v2 across all measured dimensions: higher accuracy, significantly lower completion time, and improved perceived ease of use. Notably, visually challenged participants achieved performance comparable to normal users, highlighting the system’s accessibility benefits and its potential to reduce longstanding inequities in CAPTCHA interaction.

\bigskip
\noindent\textbf{Contributions.}  
This paper makes the following contributions:

\begin{itemize}
    \item \textbf{A new CAPTCHA paradigm.} We design \textsc{ThermoCAPTC-HA}, the first human-verification system that leverages privacy preserving thermal imaging to eliminate puzzles, cognitive load, and behavioral tracking.
    \item \textbf{Farm-resistant token binding.} We introduce a traceable token mechanism that cryptographically binds a CAPTCHA result to a specific session, device context, and thermal metadata, preventing forwarding attacks that undermine existing systems.
    \item \textbf{Comprehensive real-world evaluation.} Using 286 collected thermal images, we train and deploy a YOLOv4-tiny model achieving 96.70\% accuracy with 73.60\,ms latency. We further evaluate robustness under angle variation, distance changes, camera tilt, and diverse thermal backgrounds.\footnotemark
    \item \textbf{Improved accessibility and usability.} A user study with 50 participants shows that ThermoCAPTCHA offers notably faster verification, higher correctness, and better usability—particularly for visually challenged users—compared to reCAPTCHA~v2.
\end{itemize}

\footnotetext{Appendix~\ref{DataAvailability} provides complete replication materials, including dataset, model training scripts, implementation code, and usability questionnaire.}

%-------------------------------------------------------------------------------
\section{Modern CAPTCHAs}
%-------------------------------------------------------------------------------

Modern CAPTCHA deployments are dominated by behavioral mechanisms introduced by Google in 2013~\cite{GoogleOn0:online}. Instead of relying exclusively on text distortions or object-recognition puzzles, these systems execute JavaScript in the client’s browser to observe interaction patterns—such as mouse trajectories, timing features, and device artifacts—to distinguish humans from automated bots. Contemporary behavioral CAPTCHAs generally fall into two categories: puzzle-based and puzzle-less.

\paragraph{Puzzle-based CAPTCHAs.}
Following the shift toward behavioral analysis, puzzle-based CAPTCHAs evolved from static text challenges to interactive visual tasks. Early image-recognition CAPT-CHAs, such as Google's reCAPTCHA, required users to select images containing specific objects, while Arkose Labs offered orientation puzzles and GeeTest introduced slider challenges~\cite{searles2023empirical}. reCAPTCHA~v2 combined an “I’m not a robot’’ checkbox with risk scoring derived from user behavior~\cite{reCAPTCHA35Online}. hCaptcha similarly used image classification tasks and offered dataset-labeling incentives for publishers~\cite{Accessib83:online}. Although these systems balance security and usability, they continue to pose accessibility challenges for visually challenged users and often require substantial user interaction~\cite{zhang2020ephemeral}. Moreover, visual puzzles remain vulnerable to automated solvers trained on large vision datasets.

\paragraph{Puzzle-less CAPTCHAs.}
Puzzle-less systems eliminate explicit challenges entirely and rely solely on behavioral signals to determine whether a user is human. reCAPTCHA~v2 (in its no-challenge mode), reCAPTCHA~v3, and Cloudflare’s Turnstile exemplify this trend~\cite{Announci86:online}. These services compute confidence scores based on browser metadata, device fingerprinting, and user-interaction traces collected during typical page usage, such as form submissions. Although marketed as “CAPTCHA alternatives,’’ these systems operate within the same verification flow as traditional CAPTCHAs and retain many of the same vulnerabilities—particularly their susceptibility to CAPTCHA farm outsourcing~\cite{nguyen2024c}.

\begin{figure}[tp!]
    \centering
    \Description{Diagram showing the workflow of a modern CAPTCHA.}
    \includegraphics[width=0.95\columnwidth]{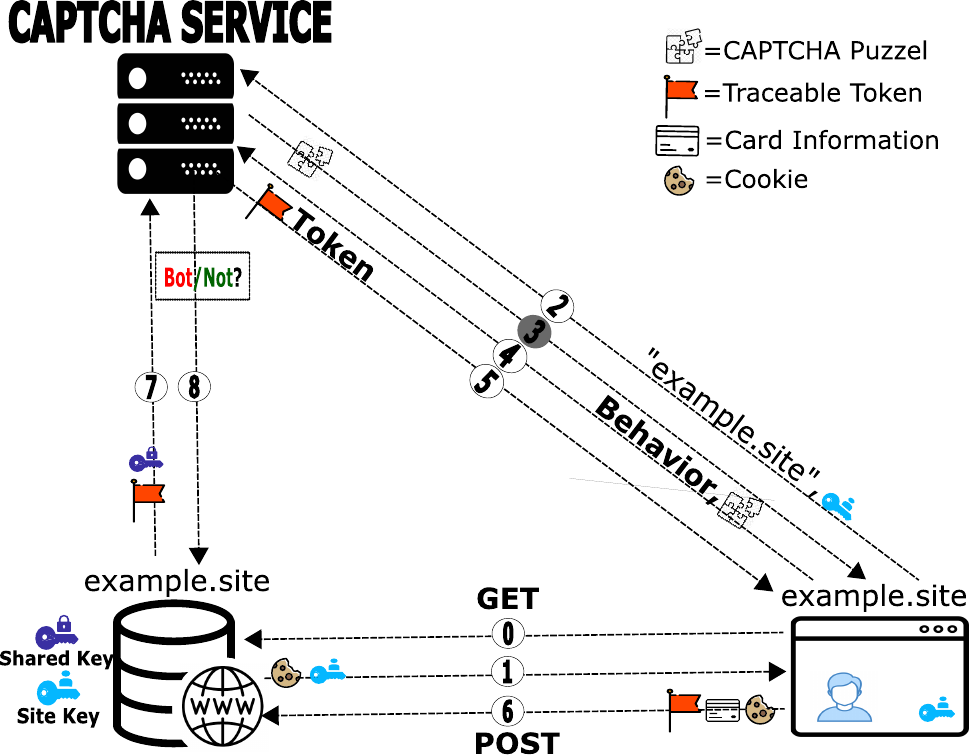}\vspace{-2mm}
    \caption{High-level workflow of a modern CAPTCHA~\cite{nguyen2024c}.}\vspace{-5mm}
    \label{fig:workingProtocolOfACAPTCHA}
\end{figure}

\subsection{Modern CAPTCHA Workflow}

Figure~\ref{fig:workingProtocolOfACAPTCHA} summarizes a typical behavioral CAPTCHA workflow. Consider a website \texttt{example.site} that deploys a CAPTCHA to protect a high-value action such as concert ticket purchases. The site embeds provider-supplied JavaScript—including a \textit{site-key} that identifies the domain—which initializes CAPTCHA collection upon page load. Providers discourage deferred or lazy loading because early execution maximizes behavioral-signal fidelity~\cite{Loadingr23:online}. The website and CAPTCHA provider share a \textit{shared-key} for secure server-to-server verification.

When a user visits \texttt{example.site}, their browser sets a cookie \circlednumblack{0} and loads the CAPTCHA script \circlednumblack{1}. The script queries the CAPTCHA provider with the \textit{site-key} \circlednumblack{2}. Since \texttt{Referer} headers may be suppressed by users or browsers, the \textit{site-key} serves as the reliable identifier for the requesting website. The provider may return an explicit puzzle \circlednumgray{3}, though puzzle-less systems omit this step. Once the puzzle is solved—or upon a relevant browser event in puzzle-less systems—the client transmits behavioral features and optional puzzle solutions back to the provider \circlednumblack{4}. The provider generates a \textit{token} encoding a risk score and returns it to the client site \circlednumblack{5}.

The website attaches this token to sensitive form data, such as payment information, and forwards it to its server \circlednumblack{6}. The server validates the token by contacting the CAPTCHA provider with the token and the site’s secret \textit{shared-key} \circlednumblack{7}. Tokens are time-limited (e.g., two minutes for reCAPTCHA) and must be validated promptly to prevent reuse~\cite{nguyen2024c}. The provider computes a final risk decision \circlednumblack{8}, which site developers interpret according to their security requirements.  

\subsection{CAPTCHA Farms}

Despite advances in behavioral modeling, modern CAPTCHAs remain vulnerable to CAPTCHA farm outsourcing~\cite{nguyen2024c}. These human-solver markets have operated for more than a decade, allowing attackers to relay CAPTCHA challenges to low-wage workers who solve them manually. Motoyama et al.~\cite{motoyama2010re} documented the economic structure of such farms, which typically consist of (1) an attacker-facing API through which CAPTCHA images or tokens are submitted, and (2) a worker interface that streams challenges to human solvers using lightweight desktop or browser tools.  

This model persists today. Services such as Anti-Captcha and Kolotibablo continue to support new behavioral CAPTCHA formats and maintain stable pricing between \$1 and \$4 per thousand solves~\cite{SolveGoo46:online,CaptchaS39:online,reCAPTCH88:online,APINewTo53:online}. Crucially, behavioral CAPTCHAs provide no mechanism to bind a verification token to the client environment that generated it. As a result, attackers can submit a CAPTCHA challenge from an automated system, relay it to a remote worker, and attach the returned token to their malicious workflow—effectively bypassing behavioral analysis and rendering puzzle complexity irrelevant. This persistent vulnerability underscores the need for CAPTCHAs that enforce cryptographic, context-sensitive token binding.

%-------------------------------------------------------------------------------
\section{Threat Model} \label{Model}
%-------------------------------------------------------------------------------

ThermoCAPTCHA operates within the standard web-security model used by existing CAPTCHA systems such as reCAPTCHA. Our threat model focuses on adversaries attempting to bypass human verification, impersonate users, manipulate thermal captures, or exploit weak token-binding mechanisms. We outline the assumptions and security boundaries below.

\paragraph{Client-Side Script Integrity and Webcam Access.}
We assume that browsers correctly retrieve and execute ThermoCAPTCHA’s Java-Script from the CAPTCHA server when the protected webpage loads. To ensure integrity, the script is delivered with Subresource Integrity (SRI) metadata; any modification to the script causes the browser to reject execution. The browser’s native permission model governs thermal-camera access, and a thermal capture is performed only when the CAPTCHA component is triggered, with the video stream terminating immediately afterward.  
\textit{We further assume that adversaries may attempt to present fabricated or manipulated thermal inputs—such as heated objects, replayed captures, or adversarially perturbed images—with the goal of deceiving the human-detection model. ThermoCAPTCHA must therefore be robust against such spoofing attempts.}

\paragraph{Server-Side Security.}
We assume a hardened and uncompromised CAPTCHA backend server. The server stores sensitive material such as nonces, session identifiers, and secret keys—encrypted at rest using AES-256 and transmits all data exclusively over secure channels (TLS~1.2 or higher). Strong access control, multi-factor authentication for administrative interfaces, periodic penetration testing, and routine key rotation are in place. The server maintains data integrity using cryptographic hash functions and enforces strict validation on all incoming client metadata and thermal-image signatures.

\paragraph{Website--CAPTCHA Server Synchronization.}
The protected website and the ThermoCAPTCHA server communicate through a pre-established \textit{shared-key}. We assume that both parties follow secure key-management practices, including regular key rotation and protection of keys at rest. Synchronization mechanisms ensure consistent timing windows for validating signed thermal metadata (e.g., timestamps, nonces), preventing acceptance of stale or replayed verification attempts. Misalignment between server and website clocks, when it occurs, is assumed to be small and bounded.

\paragraph{User Privacy Requirements.}
Thermal captures used by ThermoCAPTCHA contain coarse heat-distribution patterns that cannot uniquely identify individuals~\cite{bhowmik2011thermal,lin2021thermal,saad2023hotfoot}. Nonetheless, we assume that the service adheres to strong privacy-preserving data-handling practices. Thermal images are anonymized, encrypted at rest, and accessible only to authorized personnel under multi-factor authentication. Retention is minimized: images are stored only for the brief period needed for verification, after which they are securely deleted. These safeguards mitigate privacy risks even in the event of partial system compromise.
%%%%%%%%%%%%%%%%%%%%%%%%%%%%%%%%%%
%-------------------------------------------------------------------------------
\section{ThermoCAPTCHA System}
%-------------------------------------------------------------------------------
 
Our goal is to design a real-world, image-based CAPTCHA that (i) protects user
privacy, (ii) resists outsourcing to CAPTCHA farms, and (iii) preserves the
low-friction experience of one-click behavioral CAPTCHAs. ThermoCAPTCHA achieves
this by combining privacy-preserving thermal imaging with cryptographically
bound, farm-resistant tokens, while remaining compatible with standard web
architectures.

We distilled our design into three guiding goals:

\begin{itemize}
\item \textbf{Service-agnostic deployment.} ThermoCAPTCHA should integrate into
existing web infrastructures using the same client–server pattern as systems
like reCAPTCHA, without requiring application-specific changes or trusted
hardware, and while minimizing the attack surface exploited by CAPTCHA
farms~\cite{nguyen2024c}.

\item \textbf{Fast, web-scale operation.} Verification must complete with-in tens
of milliseconds on commodity servers to be viable for high-traffic sites and
latency-sensitive user flows.

\item \textbf{Privacy and ethical robustness.} The system should avoid identity
linkage, behavioral profiling, and continuous video capture, especially given
the ethical risks of camera-based detection systems~\cite{kristanto2023hanstreamer}.
\end{itemize}

\subsection{System Design}

ThermoCAPTCHA is designed as a drop-in replacement for traditional behavioral
CAPTCHAs. As in reCAPTCHA-style deployments, the protected website embeds a
provider-hosted JavaScript widget and uses a \emph{site-key} and \emph{shared
key} to communicate with the CAPTCHA service. Rather than collecting mouse
trajectories or displaying puzzles, the widget performs a single thermal image
capture when the protected action is triggered (e.g., form submission or
checkout), then offloads verification to the ThermoCAPTCHA server.

We initially considered alternative signals such as RGB webcam images or
keystroke/mouse dynamics to support users with visual and auditory impairments.
However, RGB images and behavioral traces raise significant privacy risks: they
enable biometric identification, cross-site behavioral profiling, and long-term
tracking even if transported over encrypted channels. These properties conflict
with our goal of minimizing data sensitivity and ethical risk. As a result, we
discarded these designs in favor of thermal imaging.

\paragraph{Thermal imaging as a privacy-preserving signal.}

Thermal cameras have lower spatial resolution than typical visible-light
cameras and primarily encode heat distribution rather than detailed texture.
Prior work shows that thermal images alone do not support reliable,
fine-grained identification of individuals, especially under realistic
variations such as masks, glasses, and changing ambient temperature
conditions~\cite{bhowmik2011thermal,lin2021thermal,saad2023hotfoot,iravantchi2024privacylens}.
External factors (e.g., activity level, emotional state, environment) further
distort thermal signatures, making them unsuitable for robust biometric
recognition.

These properties make thermal images attractive for a \emph{user-agnostic}
CAPTCHA: ThermoCAPTCHA only needs to recognize that a live human is present,
not \emph{which} human. Thermal frames therefore provide a coarse but sufficient
signal for liveness and human presence, without exposing unique identity
features. However, thermal captures are still data objects that can be
intercepted and replayed; protecting them against man-in-the-middle (MITM)
attacks requires additional cryptographic mechanisms.

\begin{figure*}[tp!]
    \centering
    \Description{Diagram showing ThermoCAPTCHA System Overview.}
    \subfigure[Protocol overview of ThermoCAPTCHA]{
    \includegraphics[width=1.10\columnwidth]{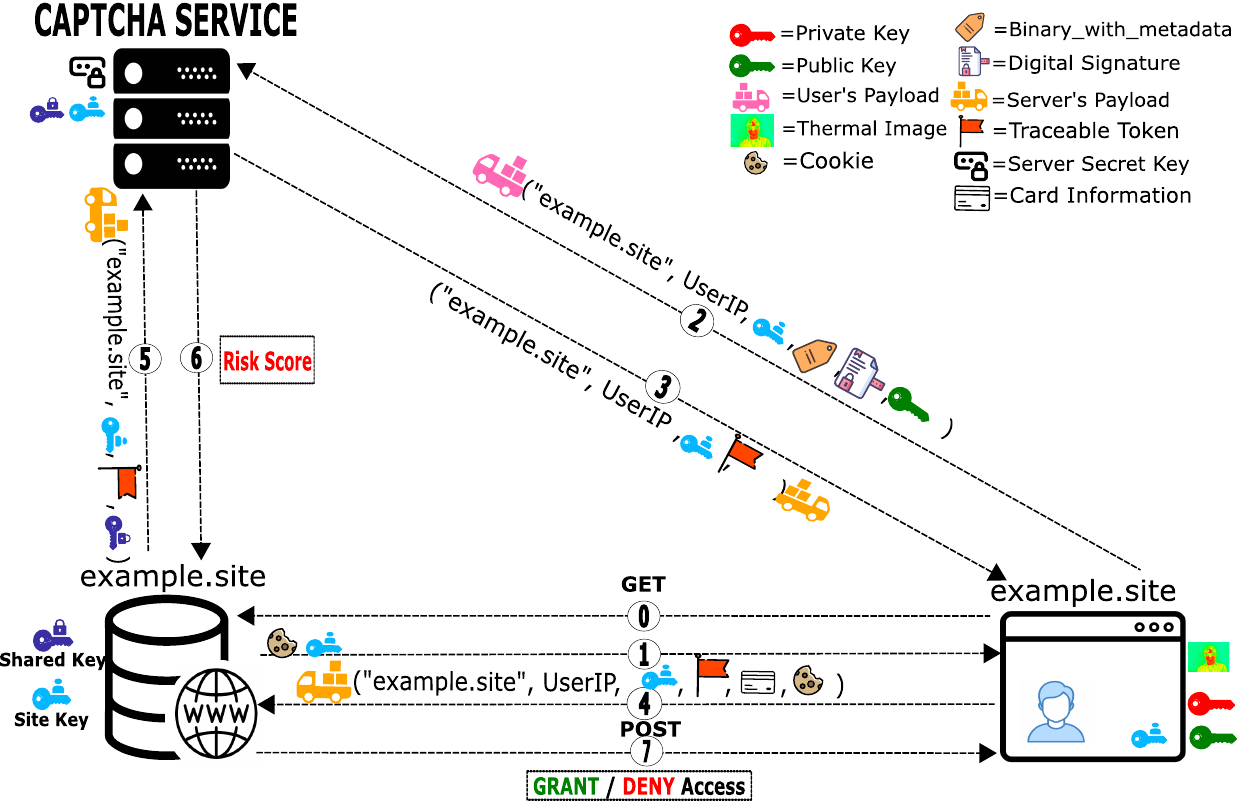}
    \label{fig:workingProtocolOfThermalCAPTCHA}
    }
    \subfigure[Digital signature process]{
    \includegraphics[width=.85\columnwidth]{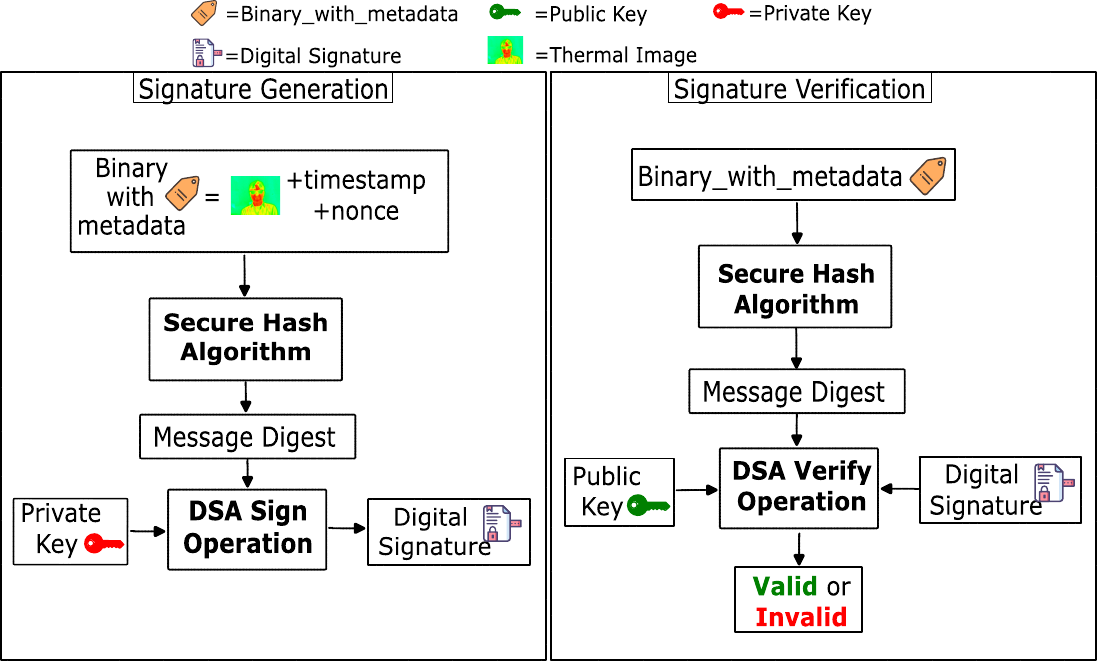}
    \label{fig:digitalSignatureProcess}
    }
    \vspace{-4mm}\caption{ThermoCAPTCHA system overview.}
    \vspace{-4mm}\label{fig:thermalCAPTCHASystemOverview}
\end{figure*}

\paragraph{Digital signatures on thermal captures.}

To defend against replay and in-transit manipulation of thermal captures,
ThermoCAPTCHA signs the hash of each captured image plus freshness metadata,
rather than encrypting the image itself. We considered two approaches:
(i) encrypting the thermal image with \texttt{AES-256} and then signing its
\texttt{SHA256} hash with \texttt{RSA}, or (ii) directly signing the
\texttt{SHA256} hash of the image and metadata using \texttt{RSA}
signatures~\cite{abdullah2017advanced,merkle1987digital}. Because thermal frames
are already privacy-preserving and do not contain unique identity features,
confidentiality is not strictly required. We therefore adopt the second
approach for lower computational overhead and simpler integration with existing
web stacks.

Concretely, when a user triggers ThermoCAPTCHA, the client widget:

\begin{enumerate}
\item Captures a thermal frame and converts it to a binary representation.
\item Appends a freshly generated \texttt{nonce} and \texttt{timestamp} to the
image metadata, yielding a \emph{binary-with-metadata} object.\footnotemark

\renewcommand{\thefootnote}{\arabic{footnote}} 
\footnotetext{Adding a nonce and timestamp to the image metadata increases the
binary size by 72 bytes without affecting human detection.}

\item Computes a \texttt{SHA256} hash of this object and signs it with the
user’s private key using \texttt{RSA}\footnotemark to produce a digital
signature.
\item Sends a payload containing the domain name, user IP, \textit{site-key},
binary-with-metadata object, digital signature, and the user’s public key to
the ThermoCAPTCHA server (Figure~\ref{fig:digitalSignatureProcess}).
\end{enumerate}

Upon receipt, the server reconstructs the hash of the binary-with-metadata
object and verifies the signature using the provided public key. Because the
hash covers the image data, nonce, and timestamp, any modification to the image
(or associated metadata) causes verification to fail. The server also enforces a
tight time window (two minutes in our implementation) and rejects reused nonces
to prevent replay of old captures. This combination of hashing and digital
signatures ensures integrity, authenticity, and freshness of the submitted
thermal frames.

\renewcommand{\thefootnote}{\arabic{footnote}} 
\footnotetext{We chose RSA over ECC due to its widespread deployment and
compatibility with existing infrastructures.}

\paragraph{Farm-resistant traceable tokens.}

CAPTCHA farms routinely break behavioral systems by solving challenges on their
own infrastructure and returning tokens to attackers via APIs
(e.g., Anti-Captcha, Kolotibablo)~\cite{SolveGoo46:online,CaptchaS39:online,reCAPTCH88:online,APINewTo53:online,nguyen2024c}.
Through MITM analysis using tools like \texttt{mitmproxy}~\cite{mitmprox49:online}, we
observe that these services forward site-specific metadata (such as the
\textit{site-key} and URL) to workers, who use browser extensions or automated
browsers to simulate visits and obtain valid tokens. Because existing CAPTCHA
tokens are not bound to a particular device or session context, attackers can
reuse them in arbitrary environments.

ThermoCAPTCHA addresses this by issuing \emph{traceable tokens} that are
cryptographically bound to the originating user session and device. For each
completed thermal verification, the server generates a unique \textit{sessionID}
and device \textit{fingerprint} and stores them in a session table. It then
constructs a payload containing the user identifier, sessionID, fingerprint,
nonce, and expiration timestamp, signs it as a JSON Web Token (JWT), and
encrypts it using Fernet with both the CAPTCHA server’s secret key and the
website’s shared key.\footnotemark{} The resulting ciphertext is returned to the
client website as the ThermoCAPTCHA token.

When the protected website later presents this token for final verification, the
ThermoCAPTCHA server decrypts it, re-checks the nonce and timestamp, and
verifies that the embedded sessionID and fingerprint match an active session
record. Any discrepancy (e.g., a token replayed from a different device or
network) causes verification to fail. As a result, tokens obtained by CAPTCHA
farm workers cannot be forwarded or reused outside their original context.

\renewcommand{\thefootnote}{\arabic{footnote}}
\footnotetext{We use Fernet for ease of integration with Python; JSON Web
Encryption (JWE) could be used equivalently.}

\subsection{Protocol Overview}

Figure~\ref{fig:workingProtocolOfThermalCAPTCHA} illustrates ThermoCAPTCHA’s
high-level protocol, whi-ch mirrors the flow of modern behavioral CAPTCHAs
(Figure~\ref{fig:workingProtocolOfACAPTCHA}) while replacing behavioral signals
with thermal captures and traceable tokens.

\begin{figure}[t]
    \centering
    \Description{Diagram showing the workflow of ThermoCAPTCHA.}
    \includegraphics[width=\columnwidth]{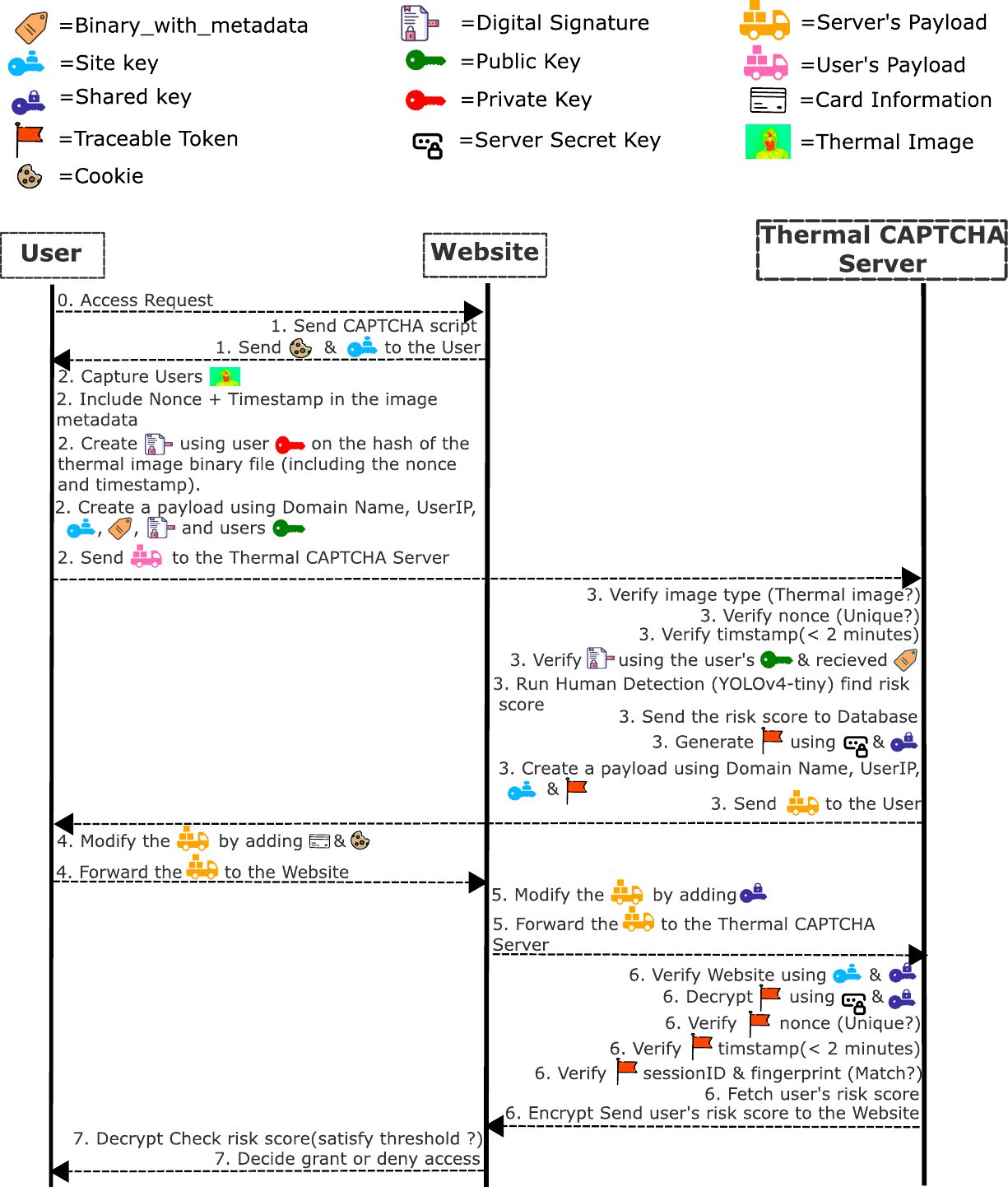}\vspace{-2mm}
    \caption{Workflow of the ThermoCAPTCHA system.}\vspace{-6mm}
\label{fig:workflowOfThermalCAPTCHASystem}
\end{figure}

At a high level, the workflow proceeds as follows
(see also Figure~\ref{fig:workflowOfThermalCAPTCHASystem}):

\begin{enumerate}
    \item \textbf{Initialization (user $\rightarrow$ website).}
    When a user visits a protected section of \texttt{example.site} (e.g.,
    ticket purchase), the website sets a cookie, loads the ThermoCAPTCHA
    script with its \textit{site-key}, and prepares to gate the critical
    action on CAPTCHA verification.

    \item \textbf{Thermal capture and signed submission (user $\rightarrow$ ThermoCAPTCHA).}
    Upon trigger (e.g., form submission), the client widget captures a thermal
    image, embeds a nonce and timestamp into its metadata, computes a hash,
    signs it with the user’s private key, and sends the resulting payload
    (including the binary-with-metadata, signature, and public key) to the
    ThermoCAPTCHA server.

    \item \textbf{Verification and token issuance (ThermoCAPTCHA $\rightarrow$ user).}
    The server validates that the payload contains a valid thermal image,
    enforces nonce uniqueness and timestamp freshness, and verifies the digital
    signature. It then runs the YOLOv4-tiny model (Section~\ref{HumanDetectionModel})
    to compute a human-detection confidence score, stores this score in its
    database, and issues a traceable token bound to the corresponding
    user/session/device context. The token is returned to the client website.

    \item \textbf{Final token validation (website $\leftrightarrow$ ThermoCAPTCHA).}
    The website attaches the ThermoCAPTCHA token to its application-level
    request (e.g., payment information) and forwards it, along with its
    shared key, to the ThermoCAPTCHA server for verification. The server
    checks the site-key/shared-key pair, decrypts and validates the token
    (including nonce, timestamp, sessionID, and fingerprint), and retrieves
    the stored risk score.

    \item \textbf{Decision (website $\rightarrow$ user).}
    ThermoCAPTCHA returns the encrypted risk score to the website, which
    decrypts it and applies local policy (e.g., thresholding) to accept or
    reject the action. The final allow/deny decision remains under the
    website’s control.
\end{enumerate}

This design preserves the deployment model of existing CAPT-CHA services while
introducing two key changes: (1) verification is based on a single thermal
frame processed by a compact detector, and (2) tokens are cryptographically
bound to the originating execution context, preventing CAPTCHA-farm relaying
without requiring changes to the web application’s control flow.

%-------------------------------------------------------------------------------
\section{Human-Detection Model (YOLOv4-tiny)} \label{HumanDetectionModel}
%-------------------------------------------------------------------------------

ThermoCAPTCHA relies on a lightweight object detector to decide whether a submitted thermal frame contains a live human. In this section, we describe the YOLOv4-tiny–based model used for this task, along with our preprocessing and training procedure.

\paragraph{Network architecture.}

Our detector is based on the \texttt{YOLOv4-tiny} architecture, chosen for its favorable trade-off between accuracy and inference speed on commodity hardware. \texttt{YOLOv4-tiny} is a compressed variant of \texttt{YOLOv4} with substantially fewer parameters and layers (29 convolutional layers and two YOLO detection heads, compared to 137 layers in the full model), enabling real-time operation on non-GPU servers. Although slightly less accurate than full \texttt{YOLOv4}, \texttt{YOLOv4-tiny} is reported to be up to eight times faster on the MS~COCO benchmark~\cite{GitHubAl48:online}, which aligns with our requirement to keep CAPTCHA verification latency in the tens-of-milliseconds range.

We adapt \texttt{YOLOv4-tiny} for binary human-presence detection in thermal images. Specifically, we configure the network as a single-class detector and adjust the convolutional filters in the layers immediately preceding each \texttt{[yolo]} detection head using the standard formula
\texttt{filters = (classes + 5) * 3} (for three anchor masks), setting \texttt{classes = 1}. The resulting model predicts bounding boxes corresponding to human heat signatures; ThermoCAPTCHA then reduces this to a binary decision (human present / not present) based on the detector outputs.

\paragraph{Image preprocessing and augmentation.}

To improve generalization from a relatively small custom dataset, we apply standard data-augmentation techniques tailored to our setting. Our base dataset consists of 286 thermal images collected from 26 participants (details in Appendix~\ref{DatasetAdditionalExperiment}). From these, we randomly select 176 images and generate 20 augmented variants per image using random geometric and photometric transformations. Following the taxonomy of Shorten and Khoshgoftaar~\cite{shorten2019survey}, we apply random rotation, translation, shear, zoom, and horizontal flipping, as well as brightness and scale adjustments.

Concretely, we set \texttt{rotation\_range = 40}, \texttt{horizontal\_flip = True}, \texttt{height\_shift\_range = 0.2}, \texttt{shear\_range = 0.2}, \texttt{zoom = 0.2}, \texttt{width\_shift\_range = 0.2}, and \texttt{fill\_mode = 'nearest'}. This process yields 3{,}520 augmented thermal images in total, exceeding the YOLOv4 guideline of at least 2{,}000 images for effective single-class training~\cite{GitHubAl48:online}. All images are resized to $416\times416$ pixels and normalized before being fed to the network.

\paragraph{Training procedure.}

We partition the augmented dataset into training, validation, and test splits to enable principled hyperparameter tuning and unbiased evaluation. Using the validation set, we perform manual hyperparameter search and obtain the following configuration: batch size of 64, input resolution $416\times416$, momentum $0.9$, weight decay $0.0005$, saturation factor $1.5$, exposure factor $1.5$, and jitter $0.3$, with an initial learning rate of $0.00261$. The loss function and optimization procedure follow the standard \texttt{YOLOv4-tiny} training setup.

Training is run for 40{,}000 iterations, after which the model reaches an average loss of $0.0509$, consistent with the range typically observed for well-converged YOLOv4-tiny detectors. We evaluate several candidate checkpoints on the held-out test set and perform a comparative analysis of the resulting weights (Appendix~\ref{DatasetAdditionalExperiment}, Table~\ref{tab:weightComparison}). The model corresponding to \texttt{yolov4-custom\_30000.weights} achieves the best mean average precision (mAP) and an overall human-detection accuracy of \textbf{96.70\%}, and is therefore selected as the deployed detector in ThermoCAPTCHA.

%-------------------------------------------------------------------------------
\section{Experimental Setup} \label{ExperimentalSetup}
%-------------------------------------------------------------------------------

\subsection{Camera Setup} \label{CameraSetup}

To implement and evaluate ThermoCAPTCHA, we constructed two thermal-imaging module configurations. The first configuration, used for dataset collection and model training, employs an OpenMV Cam H7 R2 microcontroller paired with a FLIR\circled{R} Lepton\circled{R} Adapter Module and a Teledyne FLIR Lepton~500-0771-01 thermal sensor. This combination provides high-resolution thermal frames suitable for training the YOLOv4-tiny detector.

For real-time CAPTCHA verification, we built a second configuration using the PureThermal~2 FLIR Lepton Smart I/O Module equipped with the same Lepton~500-0771-01 sensor. PureThermal~2 was chosen for its plug-and-play installation and native \textit{UVC} compatibility, enabling seamless deployment as a thermal webcam on commodity operating systems. The assembled thermal module used in our prototype is shown in Figure~\ref{fig:thermalCaptchaCameraSetup} (Appendix~\ref{ExperimentSetup}).\footnotemark

\footnotetext{Our complete plug-and-play module can be reproduced using parts from \url{https://www.sparkfun.com/products/20813}.}

\subsection{Data Collection}

Because no suitable public dataset of human thermal images exists for CAPTCHA-style liveness verification, we created a custom dataset specifically for this study. The dataset contains 286 thermal images collected from 26 participants. Each participant contributed 11 images: nine captured at 10-degree increments along the x-axis between 50° and 130°, and two additional images captured at a fixed 90° orientation with a $\pm 10^\circ$ tilt on the y-axis. This procedure ensures diversity in facial orientation and head pose, which is critical for training a robust human-presence detector.

The full experimental environment used during image collection is described in Appendix~\ref{ExperimentSetup}, and additional dataset statistics and augmentation details appear in Appendix~\ref{DatasetAdditionalExperiment}.

%-------------------------------------------------------------------------------
\section{Evaluation}
%-------------------------------------------------------------------------------
We now evaluate proposed ThermoCAPTCHA along several dimensions. Our study is guided by the following research questions:

\begin{itemize}
    \item \textbf{RQ1 — System Effectiveness.}  
    How accurately and efficiently does ThermoCAPTCHA detect human presence under realistic deployment conditions (e.g., varying angles, distances, and background environments)?

    \item \textbf{RQ2 — Security Against Adversaries.}  
    To what extent does the proposed system withstand practical attacks, including man-in-the-middle manipulation, spoofing attempts, adversarial perturbations, and cross-entity token reuse by CAPTCHA farms?

    \item \textbf{RQ3 — Robustness to Real-World Misuse and Input Manipulation.}  
    Can the system be bypassed using non-thermal images, compromised client-side execution (e.g., absence of SRI), or physical thermal forgeries such as heated objects or mannequins?
\end{itemize}

Unless otherwise stated, all experiments use the YOLOv4-tiny model described in Section~\ref{HumanDetectionModel}. All evaluations are conducted exclusively on thermal samples not used during training or validation. For consistency across experiments, a detection is considered correct if the model's confidence score exceeds 0.50.

\subsection{RQ1 — System Effectiveness}
\label{subsec:rq1}

RQ1 evaluates the practical viability of the ThermoCAPTCHA by measuring its
ability to (1) reliably detect human presence under realistic deployment
conditions, and (2) respond within latency constraints consistent with modern,
interactive CAPTCHA systems.

\subsubsection{Detection Accuracy and Latency in Controlled Conditions}

We deployed ThermoCAPTCHA on a live website backed by a Flask API and an
SQLite3 database, running on a low-power Ubuntu~20.04 server equipped with an
Intel Core i5-8550U CPU, 8~GB RAM, and an NVIDIA MX150 GPU. In a controlled
environment (\(73^\circ\)F, user distance of 3~ft), the system correctly
identified 116 of 120 thermal images, yielding an accuracy of
\textbf{96.70\%}. The end-to-end verification latency—measured from image
submission to risk-score generation—averaged \textbf{73.60\,ms}, well within
the responsiveness expected of modern CAPTCHA mechanisms. Table~\ref{tab:rq1-baseline} summarizes the baseline performance metrics.

\begin{table}[t]
\centering
\caption{Baseline effectiveness of the ThermoCAPTCHA under controlled
conditions (3~ft distance, frontal position).}\vspace{-2mm}
\label{tab:rq1-baseline}
\begin{tabular}{l c}
\toprule
\textbf{Metric} & \textbf{Value} \\
\midrule
Accuracy & 96.70\% \\
Images evaluated & 120 \\
Correct detections & 116 \\
Average latency & 73.60\,ms \\
Server hardware & i5-8550U, 8\,GB RAM, MX150 GPU \\
\bottomrule
\end{tabular}
\end{table}

\subsubsection{Impact of Viewing Angle and Camera Tilt}

We next examine robustness to natural variations in user posture and webcam
placement. Participants were recorded at horizontal angles from \(50^\circ\) to
\(130^\circ\) in \(10^\circ\) increments, with \(90^\circ\) serving as the
frontal reference. As shown in Table~\ref{tab:angle-tilt-stats}, detector
confidence peaks at \(90^\circ\) (0.91 ± 0.070) and declines toward oblique
angles, reaching 0.79 ± 0.118 at \(130^\circ\). This degradation aligns with
reduced visibility of the head and torso heat signature at extreme viewpoints.

We additionally varied the vertical tilt between \(80^\circ\), \(90^\circ\), and
\(100^\circ\). Table~\ref{tab:angle-tilt-stats} shows a modest improvement at
\(100^\circ\) (0.92 ± 0.070), which we attribute to more direct exposure of the
upper-body thermal region. These findings indicate that the system tolerates
ordinary positional variation without meaningful performance loss.

\begin{table}[t]
\centering
\caption{YOLOv4-tiny confidence across horizontal and tilt angles.
Values reported on a 0--1 confidence scale.}\vspace{-2mm}
\label{tab:angle-tilt-stats}
\begin{tabular}{@{}c c c c@{}}
\toprule
\textbf{Angle Type (°)} & \textbf{Samples (n)} & \textbf{Mean Conf.} & \textbf{Std. Dev.} \\
\midrule
\multicolumn{4}{c}{\textbf{Horizontal Viewing Angles}} \\
\midrule
50  & 10 & 0.82 & 0.095 \\
60  & 10 & 0.85 & 0.088 \\
70  & 10 & 0.88 & 0.080 \\
80  & 10 & 0.90 & 0.072 \\
90  & 10 & 0.91 & 0.070 \\
100 & 10 & 0.89 & 0.078 \\
110 & 10 & 0.86 & 0.090 \\
120 & 10 & 0.83 & 0.102 \\
130 & 10 & 0.79 & 0.118 \\
\midrule
\multicolumn{4}{c}{\textbf{Tilt Angles}} \\
\midrule
80  & 10 & 0.87 & 0.085 \\
90  & 10 & 0.89 & 0.078 \\
100 & 10 & 0.92 & 0.070 \\
\bottomrule
\end{tabular}
\end{table}

\subsubsection{Impact of User Distance}

To quantify the effect of user–camera separation, nine participants captured
thermal images at distances of 3~ft, 4~ft, 5~ft, and 6~ft. Each participant
provided three images per distance, yielding \textbf{27 samples per distance}.
Table~\ref{tab:distance-stats} shows a gradual, monotonic decline in detector
confidence as distance increases, driven primarily by reduced thermal
resolution and lower pixel density on facial features at longer ranges.
Confidence remains high at 3~ft (0.91 ± 0.042) and decreases to 0.80 ± 0.073 at
6~ft. 

Despite this reduction in confidence, the detector correctly classified all
samples as human across every distance tested. This result demonstrates that
human heat signatures remain reliably distinguishable even near the operational
limits of compact consumer-grade thermal sensors.

\begin{table}[t]
\centering
\caption{YOLOv4-tiny detection confidence at varying user--camera distances. 
Values reported on a 0--1 confidence scale.}\vspace{-2mm}
\label{tab:distance-stats}
\begin{tabular}{c c c c}
\toprule
\textbf{Distance (ft)} & \textbf{Samples (n)} & \textbf{Mean Conf.} & \textbf{Std. Dev.} \\
\midrule
3  & 27 & 0.91 & 0.042 \\
4  & 27 & 0.88 & 0.055 \\
5  & 27 & 0.84 & 0.061 \\
6  & 27 & 0.80 & 0.073 \\
\bottomrule
\end{tabular}
\end{table}

\subsubsection{Robustness to Background Variation}

\begin{figure*}[t]
\centering
\Description{Thermal examples illustrating background variation.}
\subfigure[Far-away individual]{
    \includegraphics[width=0.22\textwidth]{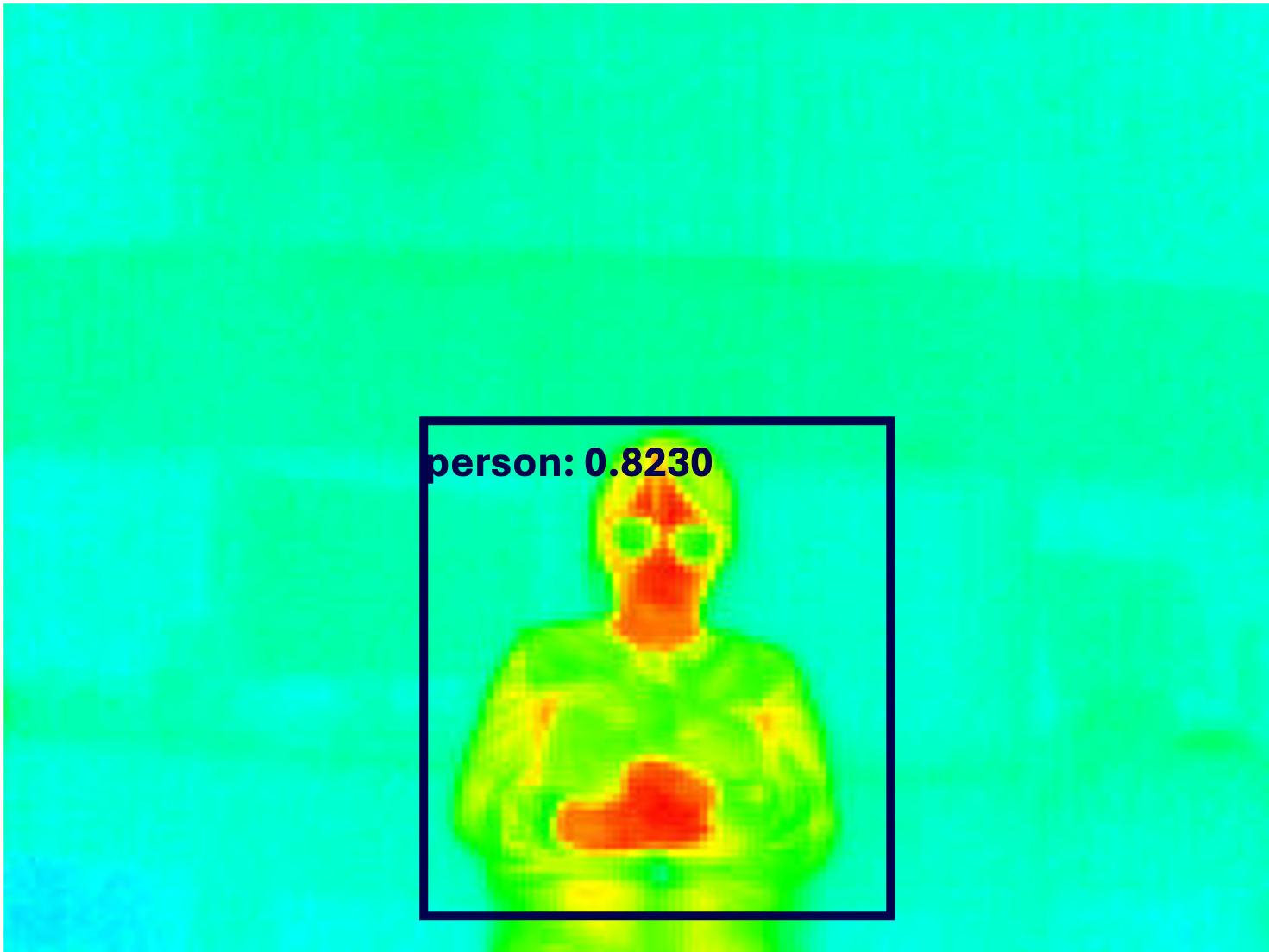}
    \label{fig:distant}
}
\hfill
\subfigure[Sunlit window]{
    \includegraphics[width=0.22\textwidth]{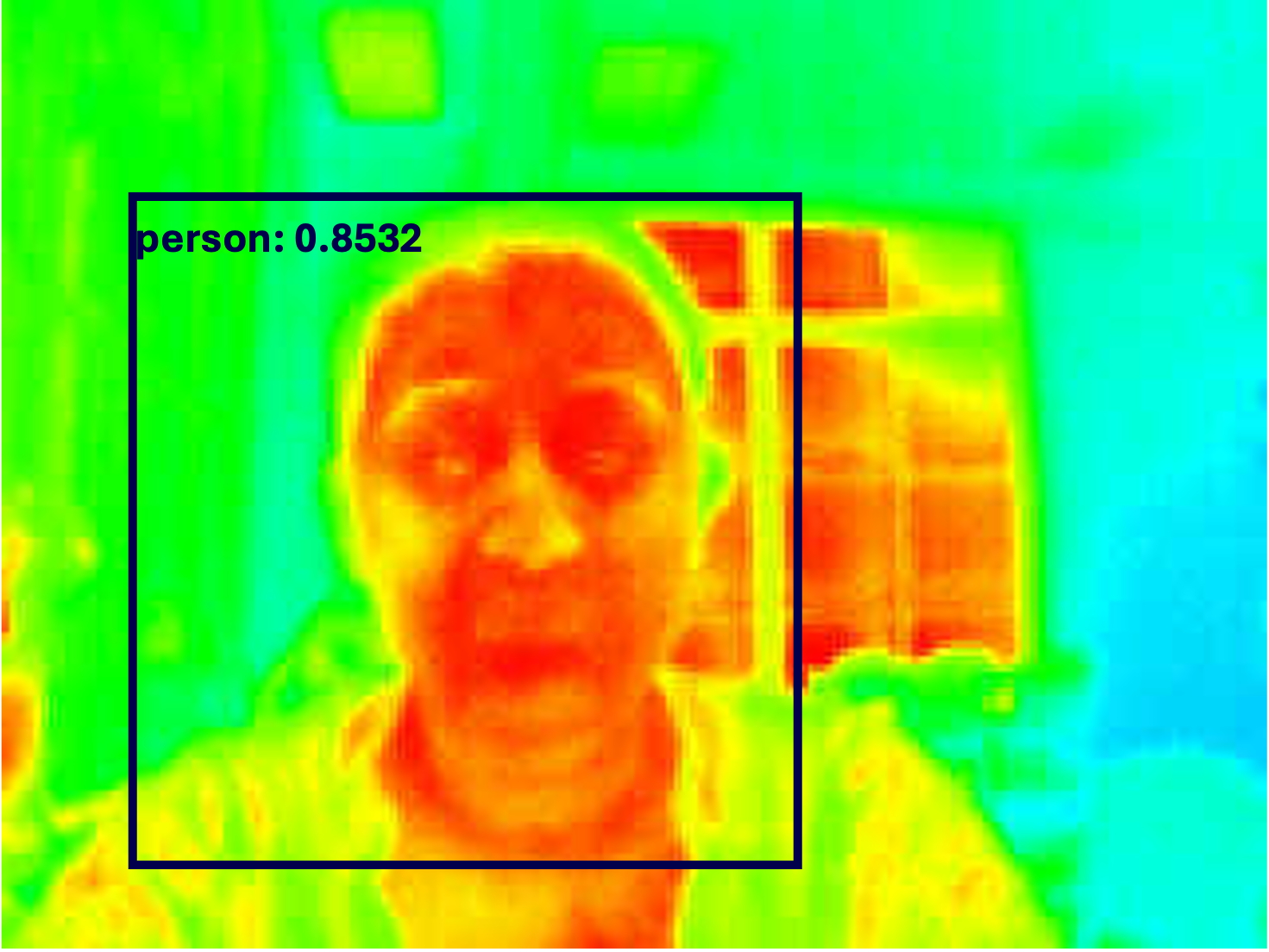}
    \label{fig:sunnyWindow}
}
\hfill
\subfigure[Heat-emitting appliance]{
    \includegraphics[width=0.22\textwidth]{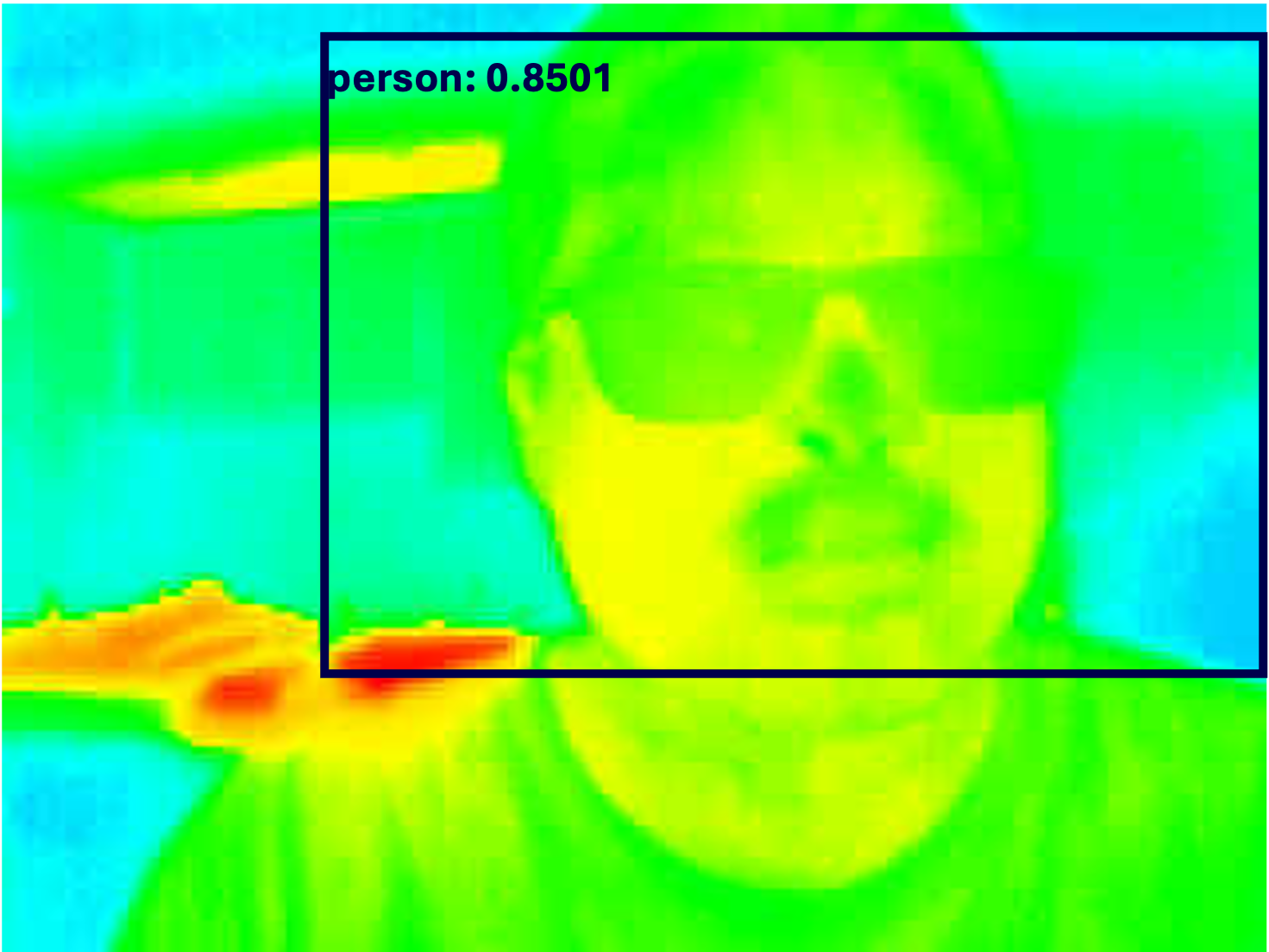}
    \label{fig:hotProduct}
}
\hfill
\subfigure[Extreme rotation ($>130^\circ$)]{
    \includegraphics[width=0.22\textwidth]{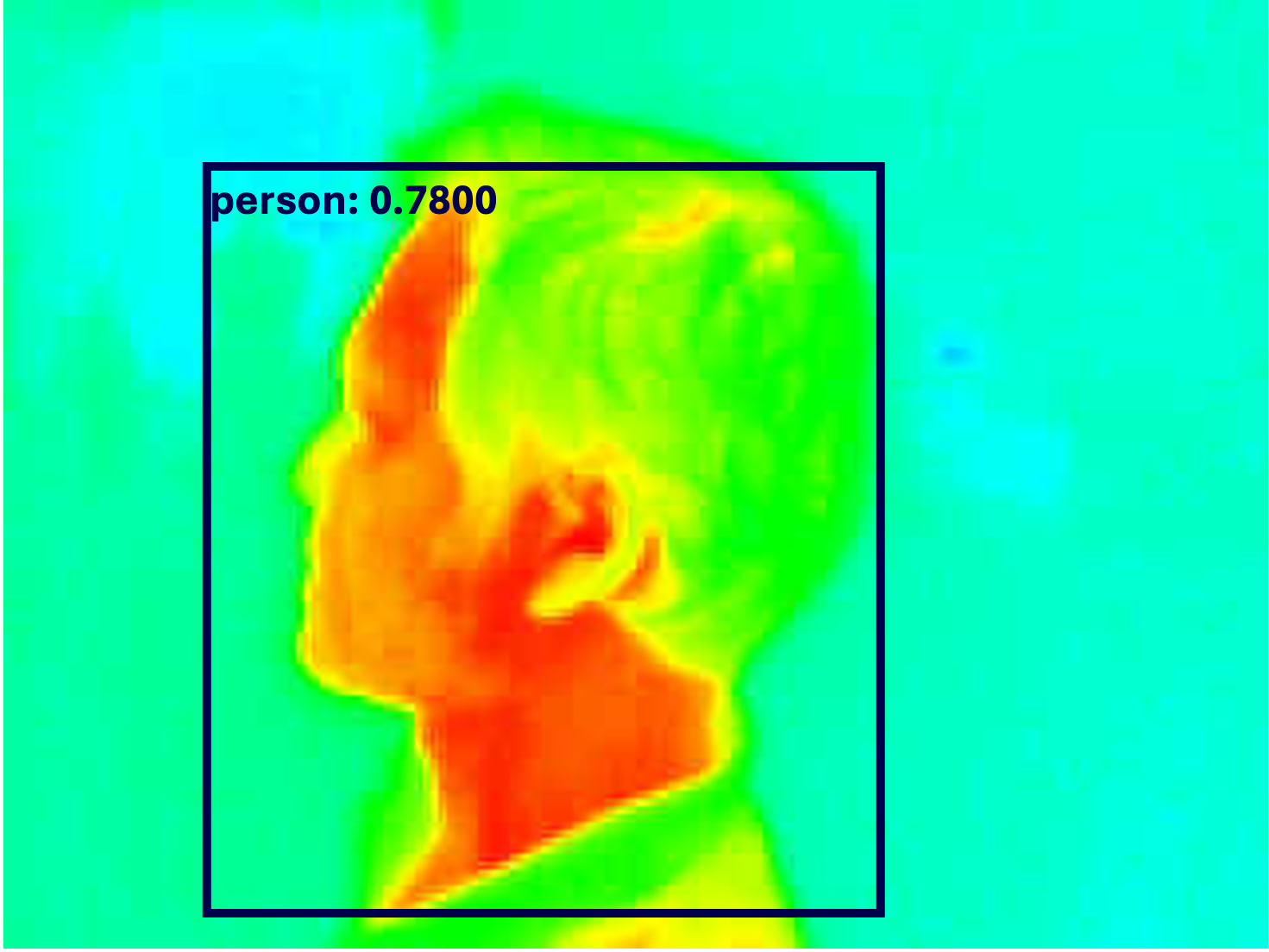}
    \label{fig:rotated}
} \vspace{-4mm}
\caption{Thermal samples collected under diverse background conditions, including 
distant subjects, direct sunlight, heat-emitting objects, and extreme body rotation. } \vspace{-4mm}
\label{fig:impactOnDifferentBackground}
\end{figure*}

To approximate real-world deployment, nine participants captured three thermal
images each in uncontrolled home environments, yielding \textbf{27 samples per
background condition}. Environments included direct sunlight, heat-emitting
appliances, low-light rooms, distant subject placement, and scenarios involving
significant body rotation. Representative thermal samples from these settings
are shown in Figure~\ref{fig:impactOnDifferentBackground}.

Table~\ref{tab:background-stats} reports the corresponding detector confidence
scores. Human detection confidence remains high (0.87–0.90) across conditions such as sunlit windows, low-light rooms, and household heat sources.
More challenging configurations—distant subjects (0.82 ± 0.071) and extreme
rotation (0.80 ± 0.075)—exhibit moderate reductions in confidence, as expected
from diminished visibility of core thermal features. Importantly, the detector
correctly classified \emph{all} samples as human, demonstrating resilience to
thermal clutter and environmental variation commonly encountered in home and
office settings.

\begin{table}[t]
\centering
\caption{Detection confidence across diverse background conditions. Each condition includes 27 thermal captures.}
\label{tab:background-stats}
\resizebox{0.97\columnwidth}{!}{
\begin{tabular}{l c c c}
\toprule
\textbf{Background Condition} & \textbf{Samples (n)} & \textbf{Mean Conf.} & \textbf{Std Dev.} \\
\midrule
Sunlit window        & 27 & 0.89 & 0.058 \\
Heat-emitting object & 27 & 0.87 & 0.062 \\
Low-light room       & 27 & 0.90 & 0.049 \\
Distant subject      & 27 & 0.82 & 0.071 \\
Extreme rotation     & 27 & 0.80 & 0.075 \\
\bottomrule
\end{tabular}
}
\end{table}

\newlength{\rqboxwidth}
\setlength{\rqboxwidth}{\dimexpr0.92\columnwidth\relax}

\fbox{%
    \parbox{\rqboxwidth}{%
        \textbf{RQ1 Findings.}
        ThermoCAPTCHA achieves high accuracy and low verification latency
        under controlled conditions, while maintaining robustness to variations
        in user posture, camera angle, distance, and background environments.
        Confidence declines predictably with user–camera distance due to sensor
        resolution limits, but all human samples were successfully detected
        across all tested scenarios. These results demonstrate that the system
        performs reliably under realistic deployment conditions.
    }%
}

\subsection{RQ2 — Security Against Adversaries}
\label{subsec:rq2}

RQ2 examines whether ThermoCAPTCHA design withstands practical attacks
from network adversaries, spoofing adversaries, and machine-learning–based
adversaries. We evaluate resilience against (1) man-in-the-middle (MITM)
manipulation, (2) token forwarding and reuse by CAPTCHA farm workers,
(3) physical and thermal spoofing, and (4) adversarial perturbations targeting
the human-detection model.

\paragraph{Adversary Model.}
We consider three classes of adversaries: (1) a network adversary capable of
observing, replaying, or modifying messages; (2) an application-layer adversary
who attempts to reuse tokens obtained through CAPTCHA-farm outsourcing; and
(3) a computational adversary who crafts spoofed or adversarial inputs to the
detection model. The adversary does not compromise the CAPTCHA server, which
is consistent with the trust assumptions of modern behavioral CAPTCHA
deployments.

\subsubsection{Resistance to Man-in-the-Middle Manipulation}

We first considered adversaries capable of intercepting, modifying, or replaying
messages exchanged between the client, web server, and CAPTCHA server. Using
\texttt{mitmproxy}, we conducted five classes of MITM attacks: (1) replay of valid
payloads, (2) submission of stale timestamps, (3) nonce reuse, (4) binary-level
tampering, and (5) metadata-only tampering. Each attack class was executed
500 times, yielding a total of 2{,}500 adversarial requests.

Table~\ref{tab:mitm-results-rq2} reports that \emph{all} adversarial attempts
were rejected, producing a 0\% success rate across all categories. Replay and
stale-timestamp attacks failed due to freshness checks, nonce reuse was blocked
by server-side uniqueness validation, and any modification to the binary or
metadata invalidated the digital signature.

\begin{table}[t]
\centering
\caption{MITM attack outcomes across 2{,}500 adversarial attempts.}
\label{tab:mitm-results-rq2}
\resizebox{0.97\columnwidth}{!}{
\begin{tabular}{l c c c c}
\toprule
\textbf{Attack Type} & \textbf{Attempts} & \textbf{Accepted} & \textbf{Rejected} & \textbf{Success (\%)} \\
\midrule
Replay            & 500 & 0 & 500 & 0.0 \\
Stale timestamp   & 500 & 0 & 500 & 0.0 \\
Nonce reuse       & 500 & 0 & 500 & 0.0 \\
Modified binary   & 500 & 0 & 500 & 0.0 \\
Modified metadata & 500 & 0 & 500 & 0.0 \\
\bottomrule
\end{tabular}
}
\end{table}

The only meaningful threat remaining is a full server compromise, which lies
outside our trust model and is similarly catastrophic for today's major CAPTCHA
deployments (e.g., reCAPTCHA, h-Captcha). These results support that the protocol enforces end-to-end integrity and
freshness properties consistent with standard cryptographic practice, while
covering the full MITM attack surface considered in prior CAPTCHA and web
security studies.

\subsubsection{Resistance to CAPTCHA-Farm Token Reuse}

A common evasion strategy for behavioral CAPTCHAs is to outsource challenges
to human solvers in CAPTCHA farms. We therefore evaluate wheth-er a valid
traceable token generated for a worker \(W\) can later be reused by an attacker
\(A\) operating under a different device fingerprint, session identifier, and IP
context.

We simulated farm-style outsourcing with \(N \in \{100, 500, 1000\}\) valid
tokens obtained by \(W\). For each token, the attacker attempted exactly one
reuse from a separate device configuration, yielding 1{,}600 total reuse
attempts. Table~\ref{tab:token-reuse-rq2} shows that none of the reuse attempts
succeeded.

\begin{table}[t]
\centering
\caption{Traceable-token reuse outcomes across 1{,}600 adversarial attempts.}
\label{tab:token-reuse-rq2}
\resizebox{0.97\columnwidth}{!}{
\begin{tabular}{l c c c c}
\toprule
\textbf{$N$ Tokens} & \textbf{Attempts} & \textbf{Accepted} &
\textbf{Rejected} & \textbf{Success (\%)} \\
\midrule
100  & 100  & 0 & 100  & 0.0 \\
500  & 500  & 0 & 500  & 0.0 \\
1000 & 1000 & 0 & 1000 & 0.0 \\
\bottomrule
\end{tabular}
}
\end{table}

These results demonstrate that forwarding a traceable token is ineffective:
the dual-key–encrypted token is cryptographically bound to the originating
\texttt{sessionID}, device fingerprint, nonce, and expiration timestamp, and
thus cannot be validated in any foreign execution context.  
This empirical finding is consistent with our formal analysis in
Appendix~\ref{Proof}, which shows that any replay attempt by an attacker \(A\)
fails the MAC verification whenever the tuple
\((UID, sess_{id}, dev_{fp}, nonce, exp)\) differs from that of the original
solver \(W\).

In contrast to behavioral CAPTCHAs such as reCAPTCHA and hCaptcha—which have
no mechanism to bind a token to a particular client—our design introduces a
cryptographically enforced, per-session binding that prevents the economic
outsourcing attack fundamental to CAPTCHA-farm operations.

\subsubsection{Resistance to Physical and Thermal Spoofing}

\begin{figure}[t]
\centering
\Description{Diagram showing a printed image of a human.}
\subfigure[Printed image]{
\includegraphics[width=0.225\textwidth]{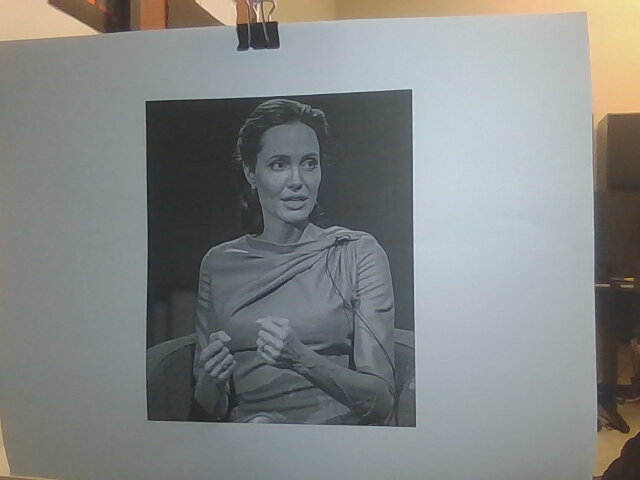}
}
\subfigure[Printed image (thermal)]{
\includegraphics[width=.225\textwidth]{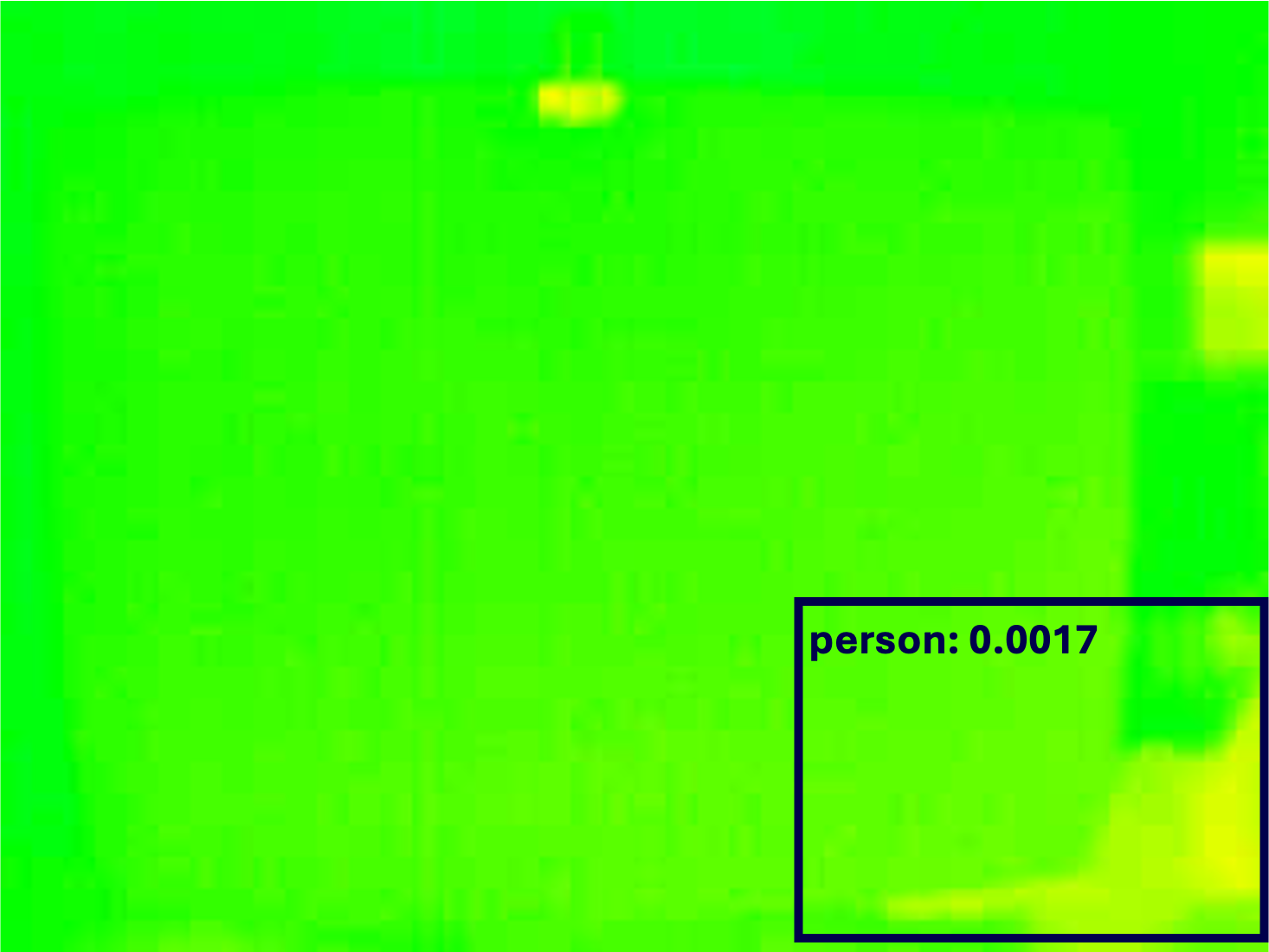}
}
\subfigure[Human mannequin]{
\includegraphics[width=0.225\textwidth]{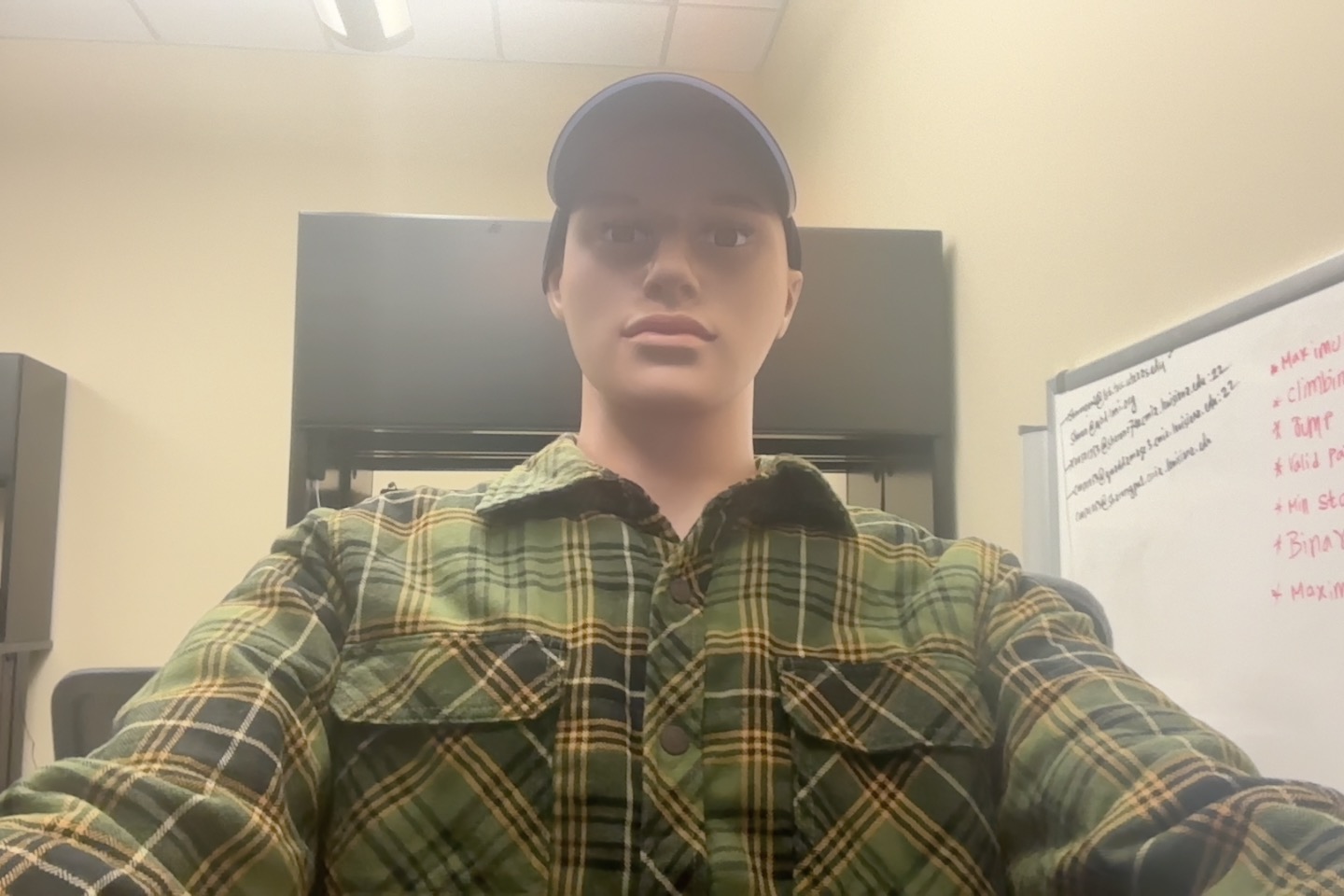}
}
\subfigure[Human mannequin (thermal)]{
\includegraphics[width=.225\textwidth]{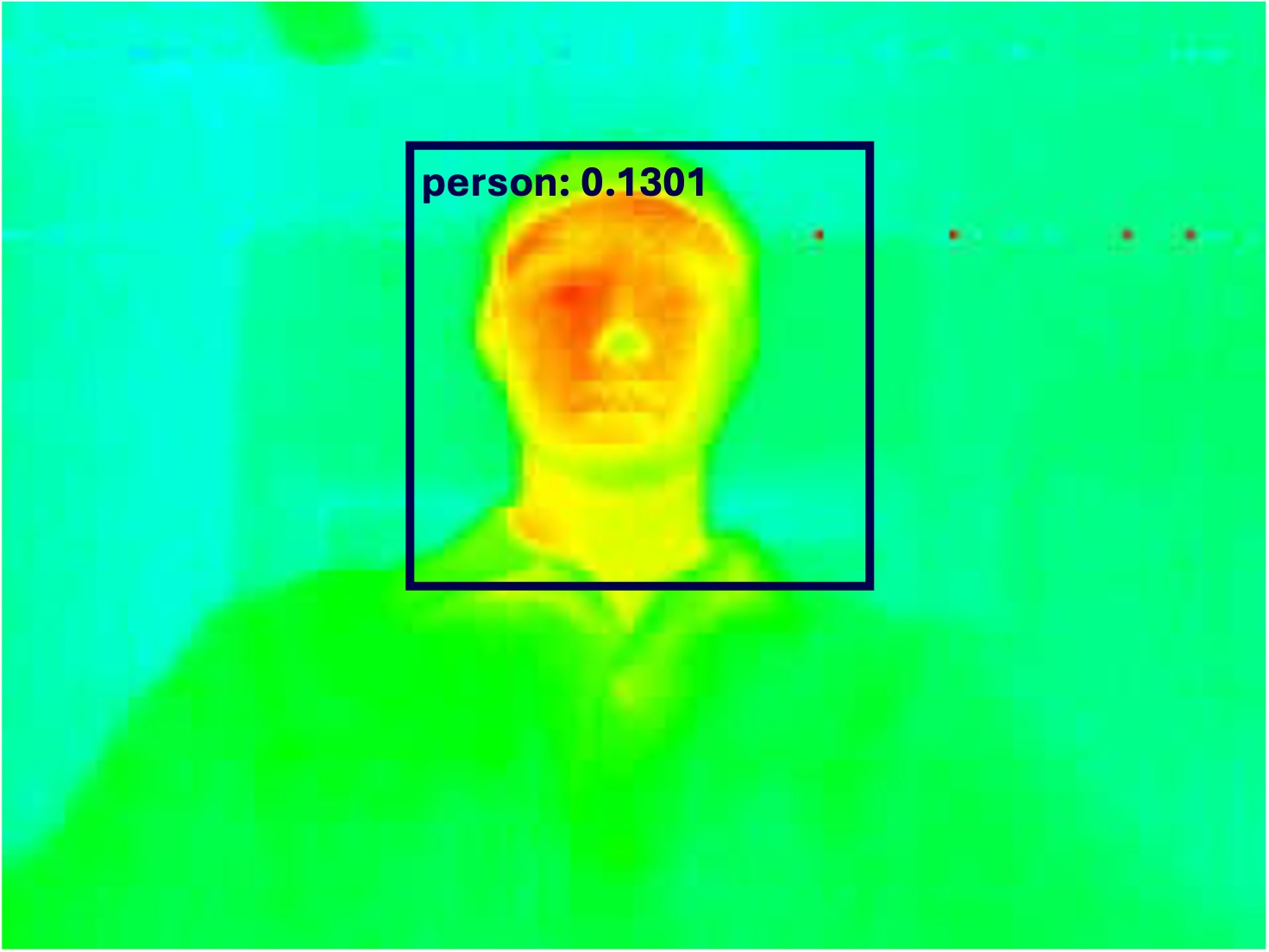}
} \vspace{-4mm}
\caption{Printed human image and heated mannequin compared with their thermal counterparts. Neither spoofed input was classified as human.} \vspace{-4mm}

\label{fig:faceSpoof}
\end{figure}

We next evaluate whether attackers can fool the system using printed photos,
thermal forgeries, or objects heated to human-like temperatures. Printed faces
(Figure~\ref{fig:faceSpoof}) produced no visible features in thermal space,
preventing the model from generating bounding boxes or confidence scores.
Heating a human mannequin with a heat gun yielded diffuse and unrealistic
thermal gradients, which YOLOv4-tiny did not classify as human.

These results indicate that reproducing the tight, high-gradient thermal
patterns of a real human body using inanimate objects is difficult in practice,
providing an inherent layer of physical liveness protection absent from
standard RGB-based CAPTCHAs.

\subsubsection{Resistance to Adversarial Perturbations (FGSM)}

Finally, to evaluate vulnerability to ML-driven evasion, we applied FGSM
perturbations to 20 thermal samples (10 human, 10 non-human) using
\(\epsilon \in \{0.01, 0.02, 0.03, 0.10, 0.20, 0.30, 0.50\}\).
A concise summary of attack outcomes is shown in
Table~\ref{tab:fgsm-summary}. The complete, per-class breakdown appears in
Appendix~\ref{AppendixFGSM}.

\begin{table}[t]
\centering
\caption{Summary of FGSM adversarial outcomes across 20 samples.}
\label{tab:fgsm-summary}
\resizebox{0.95\columnwidth}{!}{
\begin{tabular}{l c c}
\toprule
\textbf{$\epsilon$} & \textbf{Misclassifications} & \textbf{Success (\%)} \\
\midrule
0.01--0.10 & 0/20 & 0\% \\
0.20      & 1/20 & 5\% \\
0.30      & 6/20 & 30\% \\
0.50      & 16/20 & 80\% \\
\bottomrule
\end{tabular}
}
\end{table}

These results show that perturbations up to \(\epsilon \le 0.10\)
produce no errors, indicating strong intrinsic robustness of thermal patterns.
Higher-magnitude perturbations (\(\epsilon \ge 0.30\)) cause failures but
also introduce visually unrealistic artifacts that are infeasible to produce
during live thermal capture. Detailed evasion/induction results are provided in
Appendix~\ref{AppendixFGSM}.

\fbox{%
    \parbox{\rqboxwidth}{%
        \textbf{RQ2 Findings.}
        The ThermoCAPTCHA withstands practical network, spoofing, and
        ML-driven attacks. MITM manipulation and token forwarding were entirely
        unsuccessful across thousands of adversarial attempts. Physical
        spoofing attempts failed due to the unique thermal characteristics
        of the human body, while FGSM perturbations only succeed under
        unrealistically large distortion budgets. These results demonstrate
        that the proposed design provides strong protection against adversaries
        that routinely defeat conventional behavioral CAPTCHAs.
    }%
}
\subsection{RQ3 — Robustness to Real-World Misuse and Input Manipulation}
\label{subsec:rq3}

RQ3 evaluates whether the ThermoCAPTCHA fails safely under deployment
mistakes, malformed inputs, or benign but atypical usage conditions. Unlike
RQ2—which considers deliberate adversarial manipulation—RQ3 focuses on input
conditions that arise unintentionally in practice. We study three common
categories of misuse: (1) non-thermal or improperly encoded uploads, 
(2) submissions originating from compromised clients lacking Subresource 
Integrity (SRI), and (3) genuine thermal images that contain non-human heat 
sources. The non-thermal evaluation includes 80 samples across four image 
categories (20 each), while the compromised-client and non-human thermal
evaluations each use 40 and 80 samples, respectively.

\subsubsection{Handling Non-Thermal Inputs and Encoding Misuse}

To assess robustness against incorrectly formatted inputs, we submitted
80 non-thermal images to the CAPTCHA endpoint, consisting of 
20 RGB photographs, 20 grayscale images, 20 synthetic images, and 
20 desktop screenshots. Although these submissions preserved the expected
request structure, none complied with the thermal encoding required by the
server.

Across all 80 submissions, the system rejected the input during the initial
format- and metadata-validation phase, resulting in a rejection rate of
\textbf{100\%}. No malformed input propagated to the YOLOv4-tiny detector.
This quantitatively demonstrates that ThermoCAPTCHA enforces strict
type and encoding checks, ensuring that deployment errors or client misuse
fail safely and cannot be exploited as unintended authentication pathways.

\subsubsection{Execution Under Compromised or Non-SRI Client Environments}

To evaluate exposure to client-side compromise, we simulated a browser
environment without SRI enforcement. An attacker-controlled script submitted
40 manipulated payloads: 20 recompressed or modified thermal images and
20 metadata-tampered submissions (e.g., altered timestamps, missing fields, or
invalid signatures).

All 40 submissions were rejected during digital-signature validation or nonce
verification, resulting in a \textbf{0\% acceptance rate}. Even minor
re-encoding of thermal frames caused signature mismatch, indicating that the
security of the ThermoCAPTCHA does not depend on client-side integrity
controls such as SRI.

\subsubsection{Non-Human Thermal Sources and Incidental Hot Objects}

To assess robustness against benign but atypical thermal inputs, we evaluated
80 genuine thermal images that did not contain human subjects. These consisted
of 20 vacuum robot captures, 20 hot objects (e.g., a recently boiled cup),
20 cold objects, and 20 pet/animal samples. All inputs follow the correct
thermal encoding and therefore represent realistic misuse scenarios rather than
malformed data.

\begin{figure*}[t]
\centering
\Description{Thermal images illustrating non-human hot objects.}
\subfigure[Vacuum robot]{
    \includegraphics[width=0.22\textwidth]{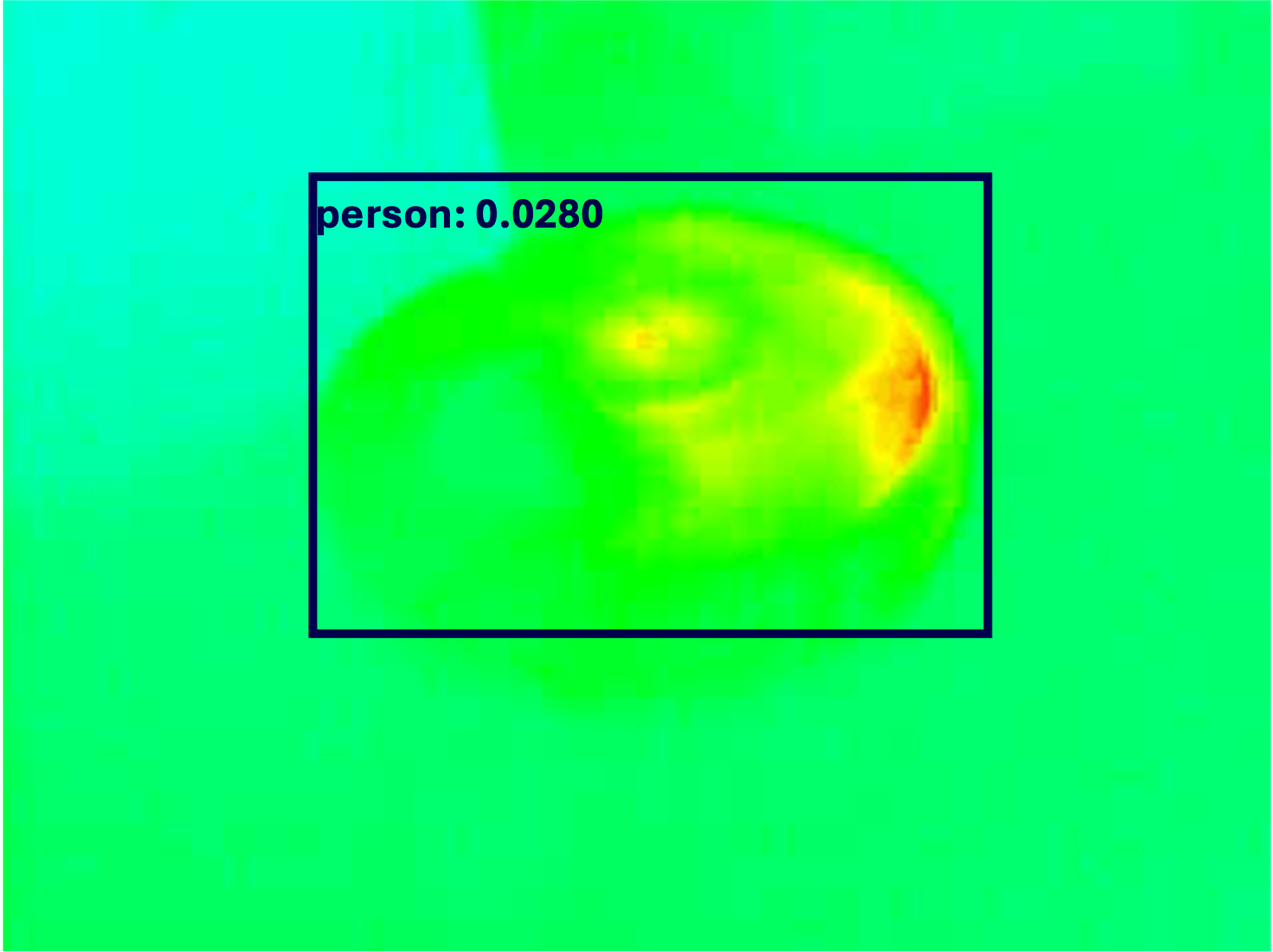}
}
\hfill
\subfigure[Hot beverage]{
    \includegraphics[width=0.22\textwidth]{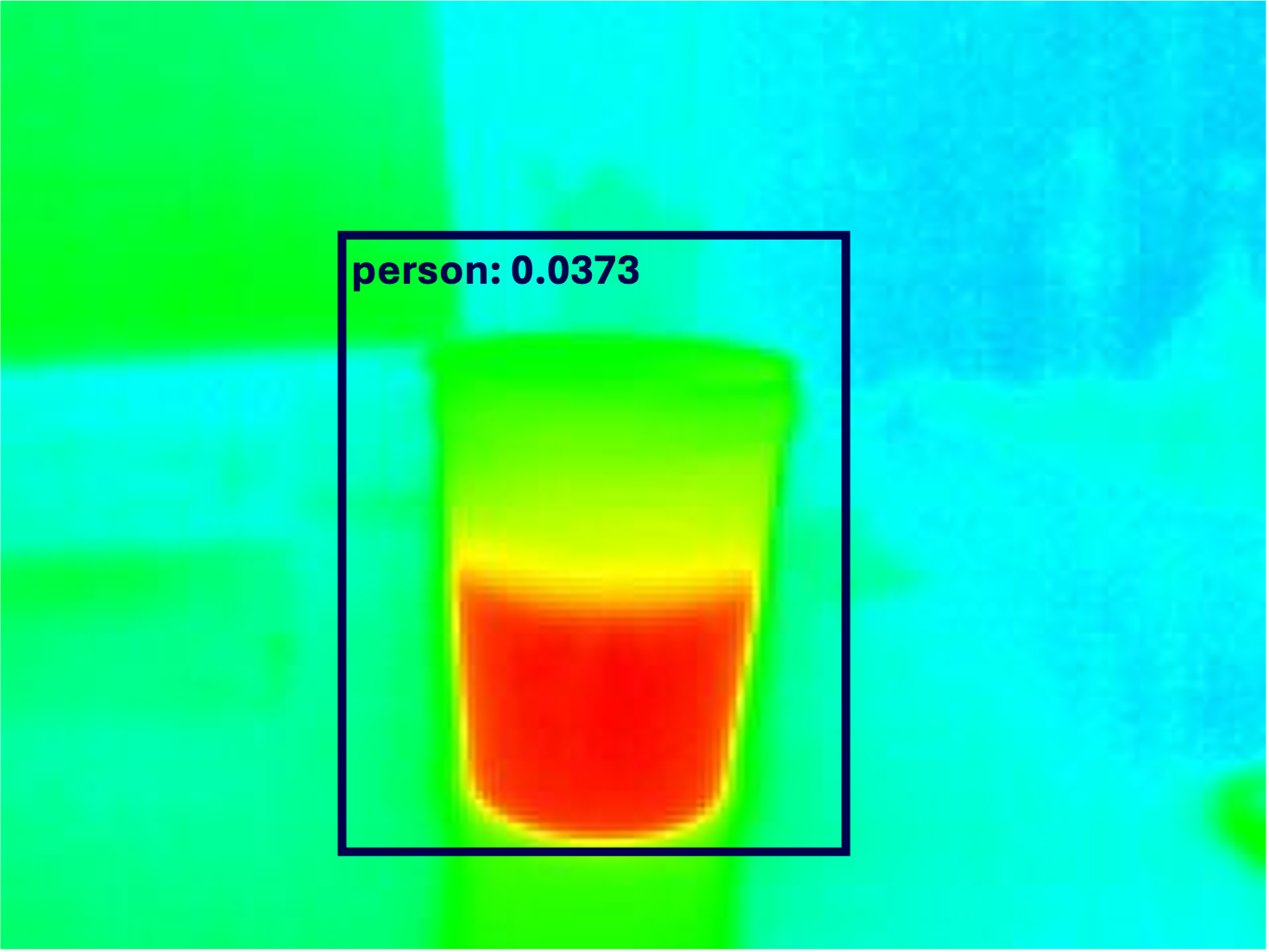}
}
\hfill
\subfigure[Cold beverage]{
    \includegraphics[width=0.22\textwidth]{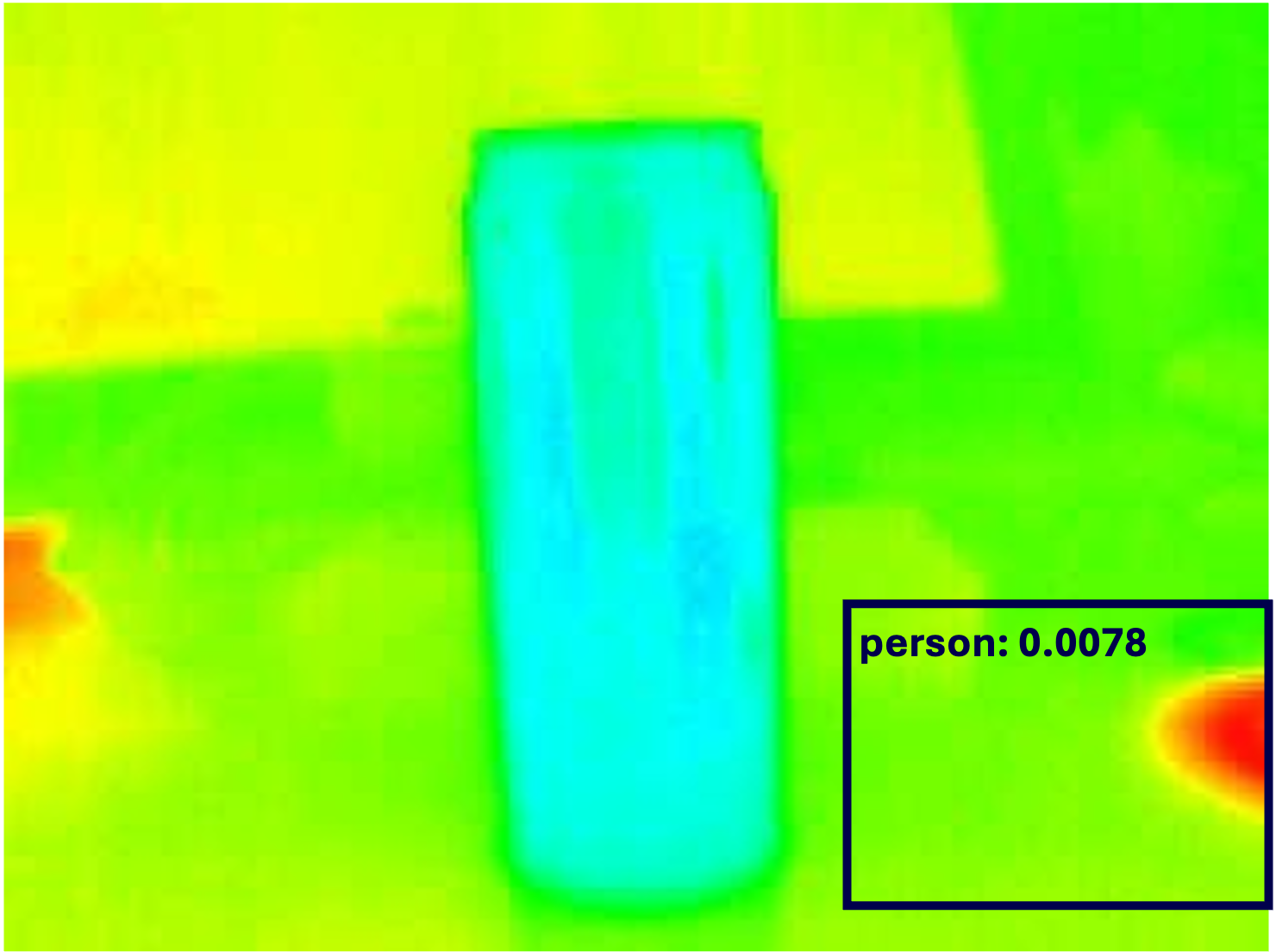}
}
\hfill
\subfigure[Cat]{
    \includegraphics[width=0.22\textwidth]{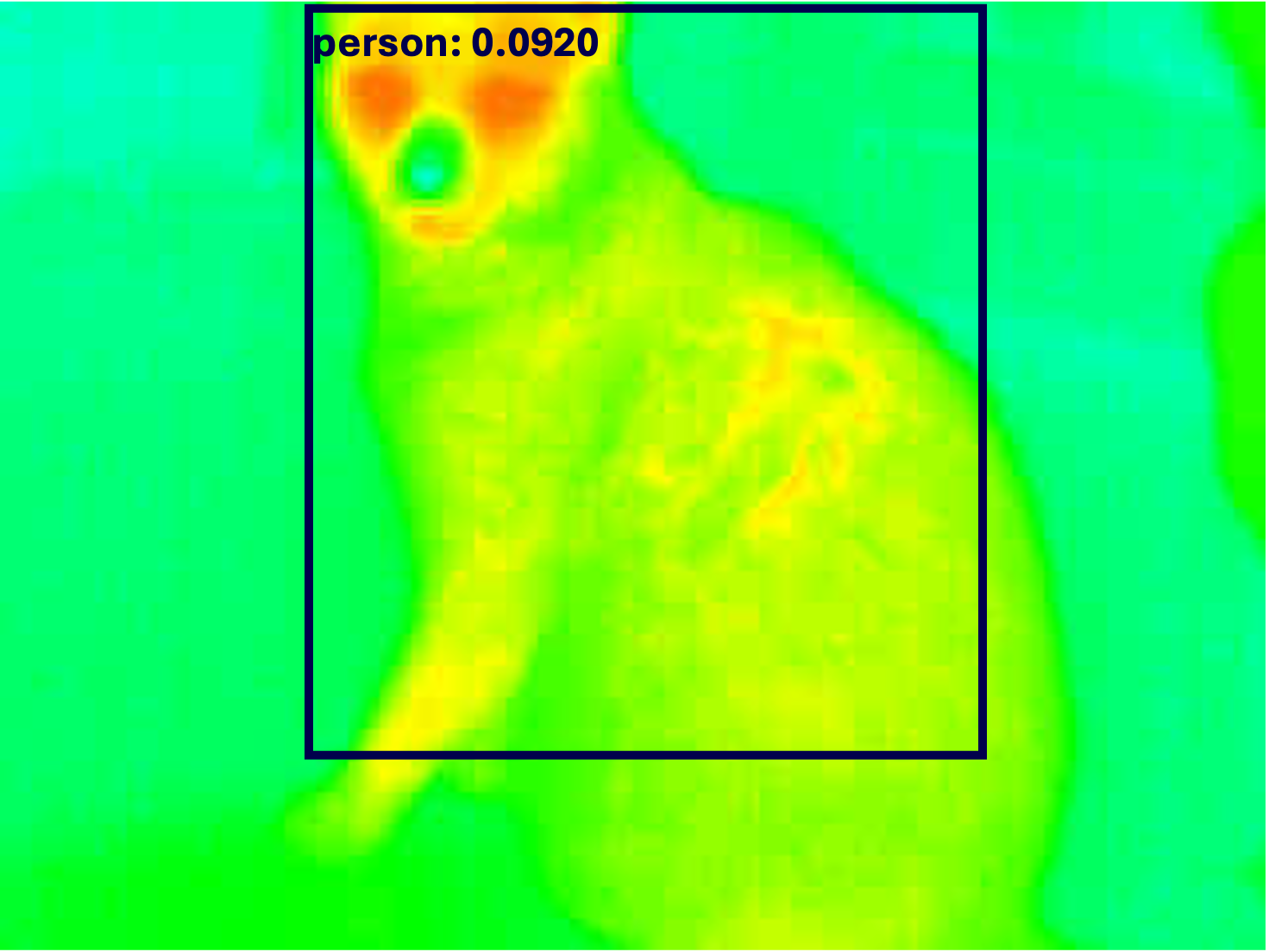}
}
\vspace{-4mm}
\caption{Examples of non-human thermal sources used to evaluate robustness to
misuse and incidental hot objects. None of these inputs was detected as human.} \vspace{-4mm}
\label{fig:impactOnOutsideTEE}
\end{figure*}

Table~\ref{tab:rq3-nonhuman} reports the corresponding detection outcomes.
Across all 80 samples, the detector produced either no bounding box or
extremely low confidence scores (\(< 0.10\)), yielding a \textbf{0\% false-acceptance
rate}. These results indicate that non-human heat signatures—even when warm,
structured, or visually similar to a human silhouette—do not mimic the spatial
and geometric thermal patterns that the detector associates with human
presence.

\begin{table}[t]
\centering
\caption{Detection outcomes for non-human thermal sources (80 samples total).}
\label{tab:rq3-nonhuman}
\resizebox{0.97\columnwidth}{!}{
\begin{tabular}{l c c c}
\toprule
\textbf{Input Type} & \textbf{Samples} & \textbf{False Accepts} & \textbf{FAR (\%)} \\
\midrule
Vacuum robot        & 20 & 0 & 0.0 \\
Hot object          & 20 & 0 & 0.0 \\
Cold object         & 20 & 0 & 0.0 \\
Pet/animal          & 20 & 0 & 0.0 \\
\midrule
\textbf{Total}      & 80 & 0 & \textbf{0.0} \\
\bottomrule
\end{tabular}
}
\end{table}

Representative examples are shown in Figure~\ref{fig:impactOnOutsideTEE}.
These results demonstrate that incidental heat sources do not trigger false
positives, and that the system preserves safety even when users unintentionally
capture warm objects, household devices, or pets during CAPTCHA interaction.

\fbox{%
    \parbox{\rqboxwidth}{%
        \textbf{RQ3 Findings.}
        The ThermoCAPTCHA fails safely under realistic deployment mistakes
        and benign misuse. All non-thermal images were rejected prior to model
        invocation, all manipulated submissions from compromised clients failed
        signature and freshness checks, and none of the non-human thermal
        sources were misclassified as humans. These results complement the
        adversarial robustness demonstrated in RQ2 and show that the system
        maintains strong safety and correctness guarantees under real-world
        operating conditions.
    }%
}

%-------------------------------------------------------------------------------
\section{User Study Methodology}
\label{userStudyMethodology}
%-------------------------------------------------------------------------------

To assess usability and accessibility, we conducted a comparative user study of \textsc{ThermoCAPTCHA} and Google reCAPTCHA. The study measured (1) task completion time, (2) correctness, and (3) subjective user experience. reCAPTCHA was integrated into our lab homepage using the public API, while \textsc{ThermoCAPTCHA} was deployed on the same server using our prototype implementation. For reCAPTCHA, we recorded the time from challenge appearance to verification; for \textsc{ThermoCAPTCHA}, we measured the time from thermal-image submission to the returned decision.

\subsection{Participants and Procedure}

We recruited 50 volunteers, including 20 visually challenged users. Participants came from multiple disciplines (including non-computer science fields), included both native and non-native English speakers, and comprised 12 women and 38 men. Each participant completed three \textsc{ThermoCAPTCHA} challenges and three reCAPTCHA challenges, receiving pass/fail feedback for each attempt. Completion times and correctness outcomes were logged automatically.

Because thermal-image–based CAPTCHAs are unfamiliar to most users, we provided a brief instructional page explaining how to solve each challenge. The detailed questionnaire layout and instructions appear in Appendix~\ref{usabilityMethodandStatistics}, and the full questionnaire text is provided in Appendix~\ref{DataAvailability}. To promote inclusivity, the study followed WCAG accessibility principles~\cite{WebConte28:online}, and an undergraduate assistant who was not involved in the design or implementation of \textsc{ThermoCAPTCHA} administered all sessions to reduce experimenter bias. On average, each participant spent 9.7 minutes completing the study.

In line with prior accessibility work and W3C reports, we used a short Likert-style questionnaire rather than the System Usability Scale (SUS) to better capture nuanced experiences of users with disabilities~\cite{Inaccess90:online,fanelle2020blind}. The questionnaire focused on perceived comfort, fun, and ease of use of \textsc{ThermoCAPTCHA} relative to reCAPTCHA.

\subsection{Completion Time and Accuracy}

Table~\ref{tab:averageTimeAndAccuracy} summarizes the Correct Attempt Ratio (CAR) and average completion time for both systems. \textsc{ThermoCAPTCHA} exceeds the 90\% usability benchmark suggested by Chellapilla et al.~\cite{chellapilla2005building}. Among normal users, the 30 participants were required to complete 90 \textsc{ThermoCAPTCHA} challenges in total (three per participant). Five of these attempts failed on the first try and required retries, resulting in 95 total attempts to obtain 90 successful completions. This corresponds to a CAR of $90/95 = 94.70\%$. For reCAPTCHA, 90 intended challenges required 109 total attempts, with 90 successful completions and 19 failures, yielding a CAR of 82.50\%.

For visually challenged participants, \textsc{ThermoCAPTCHA} similarly outperformed reCAPTCHA. With glasses, they achieved a CAR of 96.77\% for \textsc{ThermoCAPTCHA} versus 70.59\% for re-CAPTCHA; without glasses, the CARs were 95.23\% and 48.39\%, respectively. Completion times followed the same trend: normal users completed \textsc{ThermoCAPTCHA} in an average of 6.56\,s, compared to 13.32\,s for reCAPTCHA. visually challenged users required 6.75--7.36\,s for \textsc{ThermoCAPTCHA} and 14.50--19.95\,s for reCAPTCHA.

\begin{table}[t]
\caption{Correct Attempt Ratio (CAR) and average completion time for 
\textsc{ThermoCAPTCHA} and reCAPTCHA. 
``W'' = with glasses; ``W/O'' = without glasses.}\vspace{-2mm}
\label{tab:averageTimeAndAccuracy}
\centering
\resizebox{\columnwidth}{!}{%
\begin{tabular}{l cc cccc}
\toprule
\multirow{2}{*}{\textbf{CAPTCHA System}} 
& \multicolumn{2}{c}{\textbf{Normal Users}} 
& \multicolumn{4}{c}{\textbf{visually challenged Users}} \\
\cmidrule(lr){2-3} \cmidrule(lr){4-7}
& \textbf{\begin{tabular}[c]{@{}c@{}}CAR\\ (\%)\end{tabular}}
& \textbf{\begin{tabular}[c]{@{}c@{}}Avg.\ Time\\ (s)\end{tabular}}
& \textbf{\begin{tabular}[c]{@{}c@{}}W--CAR\\ (\%)\end{tabular}}
& \textbf{\begin{tabular}[c]{@{}c@{}}W--Time\\ (s)\end{tabular}}
& \textbf{\begin{tabular}[c]{@{}c@{}}W/O--CAR\\ (\%)\end{tabular}}
& \textbf{\begin{tabular}[c]{@{}c@{}}W/O--Time\\ (s)\end{tabular}} \\
\midrule
\textsc{ThermoCAPTCHA} 
& 94.70 & 6.56 
& 96.77 & 6.75 
& 95.23 & 7.36 \\
reCAPTCHA 
& 82.50 & 13.32 
& 70.59 & 14.50 
& 48.39 & 19.95 \\
\bottomrule
\end{tabular}
}
\end{table}

As shown in Table~\ref{tab:averageTimeAndAccuracy}, \textsc{ThermoCAPTCHA} 
achieves substantially faster completion times than reCAPTCHA across all user 
groups. Figure~\ref{fig:completionTime} (Appendix) provides the full 
distribution of completion times, confirming that the averages are 
representative and not driven by outliers. Notably, users unfamiliar with 
thermal images performed as well as those experienced with reCAPTCHA, 
underscoring the usability of \textsc{ThermoCAPTCHA}, particularly for 
visually challenged users.

\subsection{Subjective Usability Perceptions}

After completing all challenges, participants rated \textsc{ThermoCAPT-CHA} and reCAPTCHA on fun, ease of use, and overall comfort using 5-point Likert scales. The majority of normal users (73.30\%) and visually challenged users (60\%) preferred \textsc{ThermoCAPTCHA} on fun and ease-of-use ratings (scores 4--5). Ease-of-use ratings were particularly strong: 86.67\% of normal users and 75\% of visually challenged users rated \textsc{ThermoCAPTCHA} as easier or much easier to use than reCAPTCHA.

Participants also reported higher comfort with \textsc{ThermoCAPT-CHA} than with reCAPTCHA, frequently citing reduced visual strain and fewer repeated failures. When asked about adoption, all participants expressed interest in using \textsc{ThermoCAPTCHA} in practice, with normal users emphasizing convenience and ease of use and visually challenged users highlighting speed and reduced visual demand. Detailed distributions and plots for these subjective responses appear in Appendix~\ref{usabilityMethodandStatistics}.

%-------------------------------------------------------------------------------
\section{Discussion and Ethical Considerations}
%-------------------------------------------------------------------------------

\subsection{Limitations}

\paragraph{Single-Class Detection Strategy.}
\textsc{ThermoCAPTCHA} employs a single-class detector trained primarily on human thermal images along with synthetically derived outliers. While this strategy is effective for distinguishing “human’’ versus “non-human’’ inputs, it does not provide fine-grained object categorization. The reported accuracy therefore reflects the model’s ability to detect deviations from the thermal patterns characteristic of humans rather than general-purpose recognition performance. Future work may explore multi-class thermal detection or contrastive learning techniques to improve generalization.

\paragraph{Limited Evaluation with Users Having Hearing Impairments.}
Because \textsc{ThermoCAPTCHA} requires no audio perception, we expect strong usability for individuals with hearing impairments. However, formal user studies with hearing-impaired participants required specialized institutional approval and could not be completed within our study window. The results from visually challenged participants—high Correct Attempt Ratios (CAR), short completion times, and strong subjective ratings—suggest promising accessibility potential. A dedicated accessibility study remains an important direction for follow-up work.

\paragraph{Availability of Thermal Hardware.}
Thermal cameras are not yet standard on consumer devices, which poses a deployment barrier. Nonetheless, inexpensive smartphone-compatible modules (as low as \$38) are increasingly common and widely used in professional applications~\cite{MobileTh59:online}. \textsc{ThermoCAPTCHA} is designed to leverage such accessories when available, while falling back to a traditional CAPTCHA when thermal imaging is not supported. This hybrid design preserves inclusivity and avoids locking users out of protected services.

\subsection{Security and Privacy Considerations}

\textsc{ThermoCAPTCHA} strengthens defenses against automated attacks by combining thermal-based human verification with cryptographically bound, single-use traceable tokens. Unlike puzzle-based and behavioral CAPTCHAs—which are vulnerable to machine learning solvers, mimicry attacks, and CAPTCHA-farm outsourcing—our design prevents token forwarding and enforces strong freshness and integrity guarantees. Furthermore, thermal images are inherently low-resolution and non-identifying; the system processes them in real time and discards them immediately, reducing privacy exposure. Because the CAPTCHA service is isolated from the relying website, cross-site profiling risks are minimized. Future enhancements include exploring differential privacy on metadata and evaluating deployment within trusted execution environments (TEEs).

\subsection{Ethical Considerations}

We implemented procedural and technical safeguards to ensure responsible use of thermal imaging. The research protocol—including dataset collection and usability testing—was reviewed and approved by the university Institutional Review Board (IRB). The study was classified as minimal risk, and data access was restricted to a small set of authorized researchers.

Thermal images collected during the study were processed directly by the model to compute variance metrics (Section~\ref{ExperimentalSetup}). Volunteers received clear consent documentation outlining data handling, risks, and withdrawal rights. An independent undergraduate—uninvolved in system design—administered the user study to reduce experimenter bias.

For usability testing, both \textsc{ThermoCAPTCHA} and reCAPTCHA were presented in randomized order on a lab-hosted server. All participation occurred anonymously: Google Forms did not require login, sessions were conducted in Chrome’s incognito mode, and cookies were cleared between trials to avoid persistent camera permissions. Participants explicitly granted temporary thermal-camera access according to browser policy. These measures ensured transparency, privacy, and fair evaluation across all users.

%-------------------------------------------------------------------------------
\section{Related Work}
%-------------------------------------------------------------------------------

\paragraph{Thermal Imaging for Security and Privacy.}
Thermal cameras have been used in prior research to study unintended information leakage from user interactions. Abdelrahman et al.\ demonstrated that thermal residues left on touchscreens can allow adversaries to infer PINs and unlock patterns~\cite{abdelrahman2017stay}. Subsequent work examined the persistence of thermal traces and factors influencing attack success~\cite{Research17:online, senalp2022effects, george2019biometric}. These studies illustrate the sensitivity of thermal signals for reconstructing user actions. In contrast to these attack-focused applications, \textsc{ThermoCAPTCHA} leverages thermal imaging defensively: instead of extracting sensitive information, it relies on coarse heat distributions to verify human presence while preserving anonymity.

\paragraph{Facial Authentication and Liveness Detection.}
Modern smartpho-nes frequently employ facial recognition systems~\cite{fathy2015face, mahbub2016partial}, though many still rely on 2D imagery, which is vulnerable to replay or print attacks. Farrukh et al.\ proposed FaceRevelio, a liveness-detection approach that reconstructs coarse 3D facial geometry using a single RGB camera illuminated by dynamic light patterns~\cite{farrukh2020facerevelio}. Similar techniques such as SciFi use randomized illumination patterns for stronger spoof resistance~\cite{xie2021scifi}. However, these methods depend heavily on controlled lighting and are sensitive to strong ambient light or occlusion. In contrast, \textsc{ThermoCAPTCHA} uses thermal imaging, which is robust to illumination changes and focuses only on the presence of a heat-emitting human body rather than identity recognition.

\paragraph{Interaction-Based CAPTCHAs.}
Many traditional CAPTCHA sch-emes require explicit user interaction, such as clicking, dragging, or locating target regions. SACaptcha prompts users to identify specific shapes in an image~\cite{tang2018research}, VAPTCHA requires tracing a randomly generated trajectory~\cite{jingxia2020variation}, and Ali et al.\ introduced a drag-and-drop puzzle CAPTCHA~\cite{ali2014development}. While these approaches can resist simple bots, they impose dexterity requirements and raise accessibility concerns for users with motor impairments or limited device precision. \textsc{ThermoCAPTCHA} removes these interaction burdens, requiring only a single capture event rather than manual input.

\paragraph{Cognitive and Behavioral CAPTCHAs.}
A number of recent systems employ cognitive challenges or behavioral biometrics. Gametrics uses mouse-movement patterns in a dynamic task~\cite{mohamed2016gametrics}; Tencent, GEETest, and Netease deploy sliding-image CAPTCHAs requiring puzzle-piece alignment~\cite{zhao2018towards}; EYE-CAPTCHA relies on users solving mathematical tasks via eye movement tracking~\cite{siripitakchai2017eye}; and Mantri et al.\ use motion-based interactions from accelerometer data~\cite{mantri2018user}. Guerar et al.\ proposed an ``Invisible'' CAPTCHA that evaluates user behavior without explicit challenges~\cite{guerar2018invisible}. However, many of these systems suffer from usability issues, particularly for older adults and users with cognitive or physical impairments. Moreover, Zhao et al.\ showed that sliding CAPTCHAs can be solved with up to 98\% success using automated attacks~\cite{zhao2018towards}. 

\textsc{ThermoCAPTCHA} addresses these limitations by avoiding inter-action-heavy or cognitively demanding tasks and by employing thermal features that are difficult to spoof while requiring minimal effort from the user.

%-------------------------------------------------------------------------------
\section{Conclusion}
%-------------------------------------------------------------------------------
This paper introduced \textsc{ThermoCAPTCHA}, a thermal-image–based human-verification mechanism that combines privacy-preserving sensing with cryptographically bound, single-use traceable tokens to counter CAPTCHA-farm forwarding. Our design addresses key limitations of existing CAPTCHA systems, which often impose cognitive or visual burdens and remain vulnerable to token reuse.

Our empirical evaluation shows that \textsc{ThermoCAPTCHA} achieves 96.70\% human-detection accuracy with low verification latency, remains robust across realistic variations in angle, distance, and background conditions, and withstands MITM, spoofing, and adversarial attacks. A 50-participant usability study further demonstrates faster completion times and higher success rates than reCAPTCHA~v2, particularly for visually challenged users.

While thermal sensors are not yet standard on consumer devices, their growing availability suggests a viable path toward broader adoption. Future work includes expanding cross-environment datase-ts, integrating differential privacy, and exploring deployment within trusted execution environments. Overall, \textsc{ThermoCAPTCHA} offers a promising direction for next-generation CAPTCHAs that balance usability, accessibility, and security.

% \begin{acks}
% \end{acks}

\bibliographystyle{ACM-Reference-Format}
\bibliography{thermalCaptcha}

\appendix

%%%%%%%%%%%%%%
%-------------------------------------------------------------------------------
% Camera Modules and Experimental Setup Figures
%-------------------------------------------------------------------------------

\begin{figure}[!htp]
\centering
\Description{Thermal camera modules used for dataset collection and for the USB thermal webcam configuration.}
\subfigure[Module for dataset collection]{
    \includegraphics[width=0.40\textwidth]{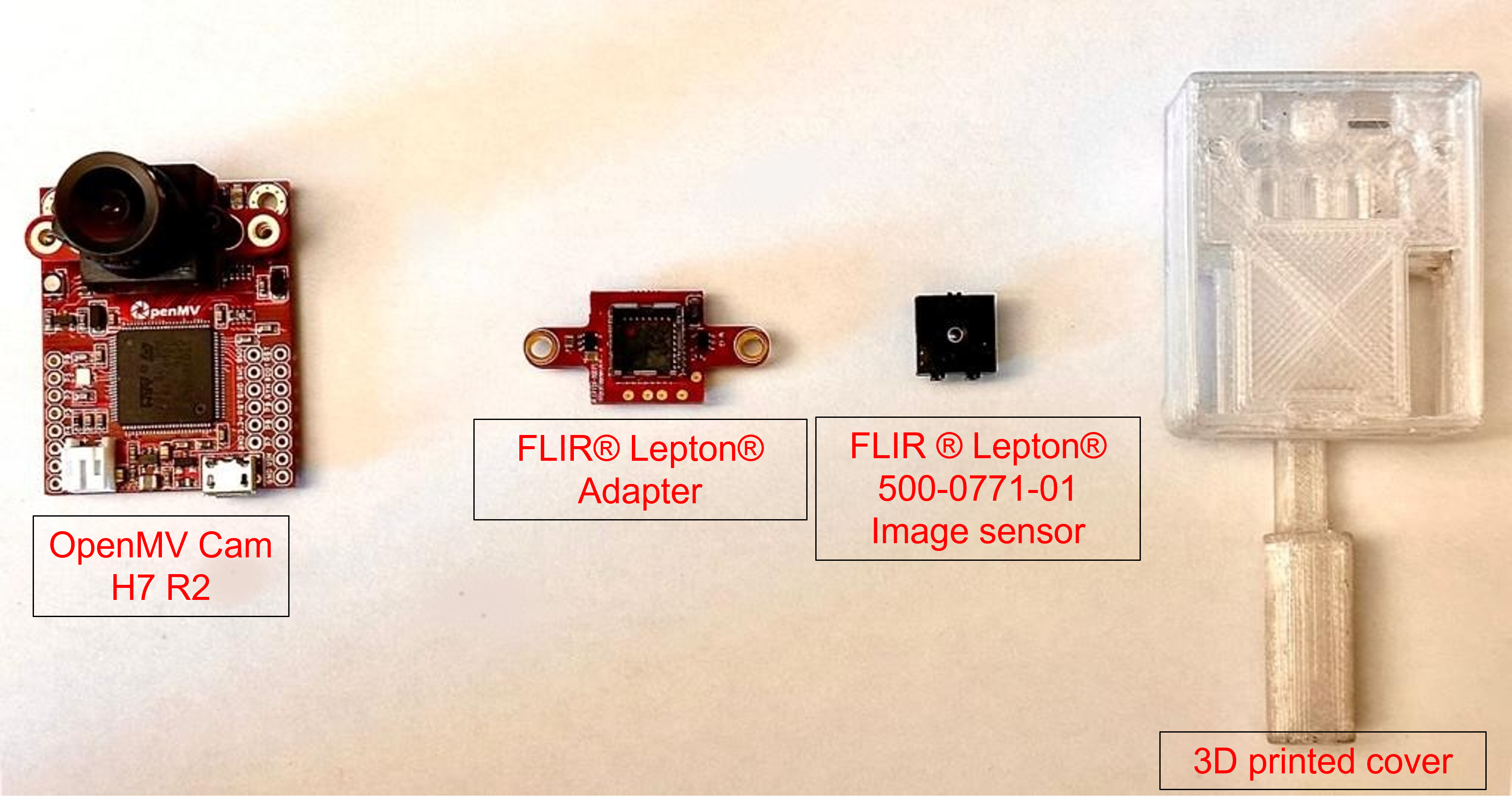}
    \label{fig:thermalCaptchaCameraSetupDataCollection}
}
\subfigure[Module for real-time webcam capture]{
    \includegraphics[width=0.40\textwidth]{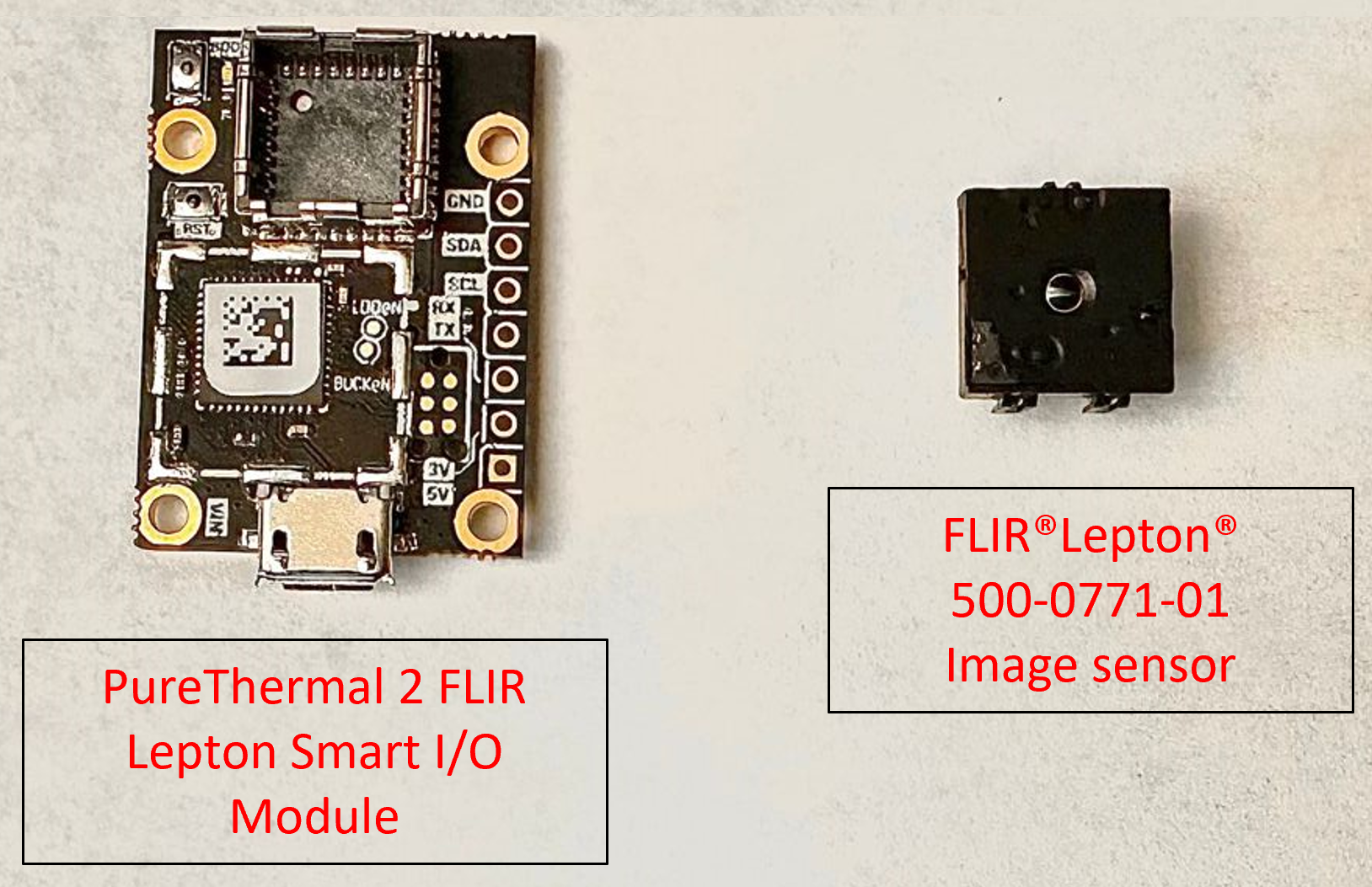}
    \label{fig:thermalCaptchaCameraSetupWebcam}
}
\caption{Thermal imaging modules used in our study.}
\label{fig:thermalCaptchaCameraSetup}
\end{figure}

\begin{figure}[!htp]
\centering
\Description{Experimental environment used for thermal data collection, including physical setup and geometric diagram.}
\subfigure[Physical data-collection setup]{
    \includegraphics[width=0.40\textwidth,height=1.8in]{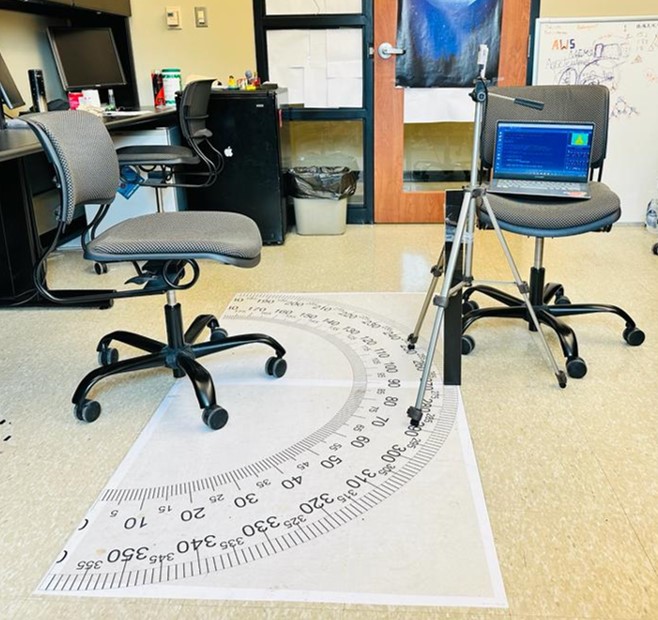}
    \label{fig:experimentalSetup1}
}
\subfigure[Schematic diagram of the setup]{
    \includegraphics[width=0.45\textwidth,height=1.8in]{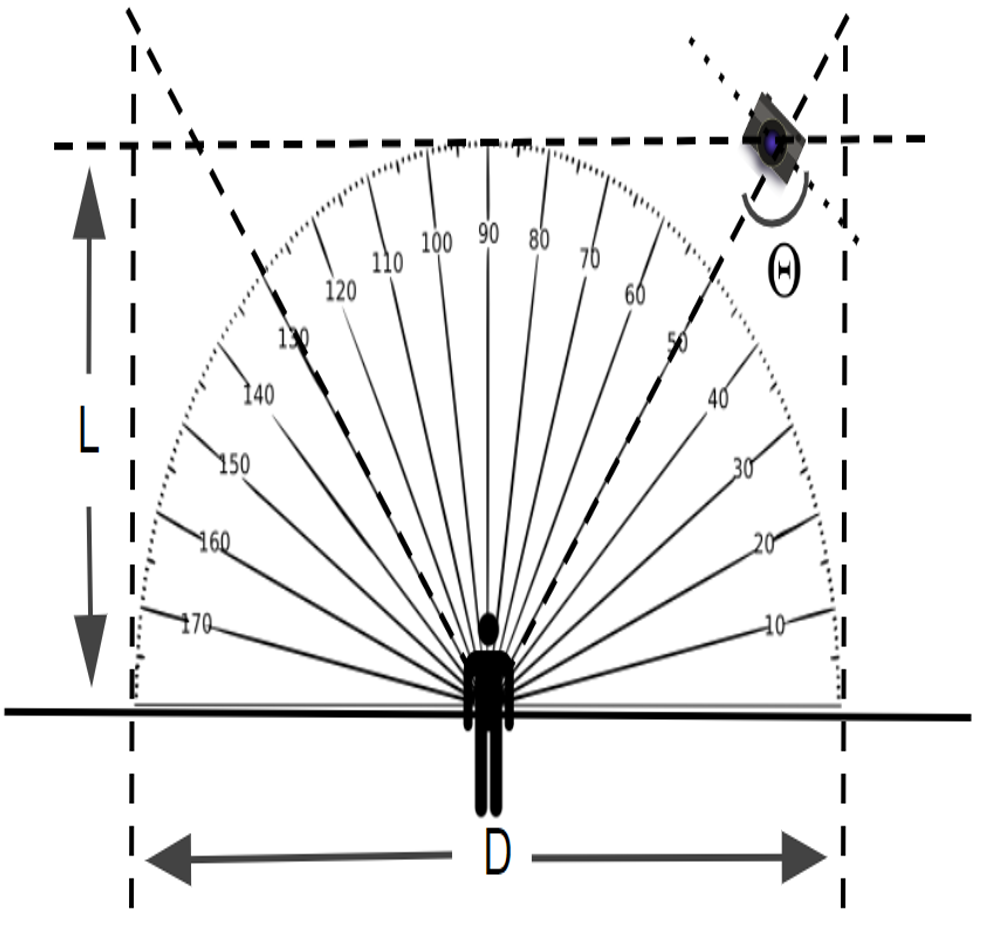}
    \label{fig:experimentalSetup2}
}
\caption{Data-collection environment. In (b), $D$ is the circle diameter defining subject positions, $L$ is the camera–subject distance, and $\Theta$ is the viewing angle.}
\label{fig:experimentalSetup}
\end{figure}

%-------------------------------------------------------------------------------
\section{Data Availability}
\label{DataAvailability}
%-------------------------------------------------------------------------------

To support replication, we provide all artefacts necessary to reproduce our
system, training pipeline, and usability study, subject to the constraints of
our Institutional Review Board (IRB) protocol. The IRB requires that access to
raw thermal-image data be restricted to researchers who have completed
appropriate human-subjects training. Accordingly, the thermal dataset will be
made available upon request to individuals who provide proof of such training
(e.g., CITI certification), consistent with the approved data-handling
procedures. An anonymous GitHub repository\footnotemark{} contains the following artefacts:

\begin{itemize}
    \item detailed training instructions, configuration files, and augmentation scripts,
    \item pre-trained YOLOv4-tiny model weights used in our evaluation,
    \item the complete ThermoCAPTCHA implementation (client widget, server backend, and database schema),
    \item and the usability-study questionnaire referenced in Section~\ref{userStudyMethodology}.
\end{itemize}

These artefacts enable end-to-end reproduction of the training pipeline,
system deployment, and usability evaluation. Researchers requesting access to
the raw thermal dataset will receive it after meeting the IRB-mandated
training requirements.

\footnotetext{Anonymous replication package:
\url{https://github.com/anonymouspaul96/Live-Thermal-Image-based-CAPTCHA}.}

%-------------------------------------------------------------------------------
\section{Data Collection Setup}
\label{ExperimentSetup}
%-------------------------------------------------------------------------------

Figures~\ref{fig:thermalCaptchaCameraSetupDataCollection} and
\ref{fig:thermalCaptchaCameraSetupWebcam} show the two thermal-camera modules used in our experiments. The first module was used to collect the dataset and train the model; the second served as a plug-and-play USB thermal webcam for real-time CAPTCHA verification.

The physical recording environment is shown in Figure~\ref{fig:experimentalSetup1}, and its corresponding geometric layout appears in Figure~\ref{fig:experimentalSetup2}. As indicated in the diagram, $D$ represents the circular arc diameter along which subjects rotated to vary horizontal viewing angle, $L$ is the fixed camera–subject distance, and $\Theta$ is the angle subtended at the camera. These figures document the controlled capture conditions employed during dataset creation.

%-------------------------------------------------------------------------------
\section{Dataset and Weight Analysis}
\label{DatasetAdditionalExperiment}
%-------------------------------------------------------------------------------

This section provides supporting details on the construction of the thermal
dataset used for model training and evaluation, as well as an analysis of the
YOLOv4-tiny weight checkpoints generated during the training process. The main
paper reports aggregate results; here we include full dataset composition and
per-weight performance metrics for transparency and replicability.

\subsection{Dataset}

A total of \texttt{286} thermal images were collected from \texttt{26}
participants, each contributing \texttt{11} images under controlled pose and
angle conditions. Of these, \texttt{110} images were allocated to the
test set, and an additional \texttt{10} images of non-human thermal objects
(e.g., hot containers, appliances) were reserved for robustness evaluations.
The remaining \texttt{176} images were used for model training.

To expand the training corpus, we applied standard geometric and photometric
augmentations, generating \texttt{20} variants per image and producing
\texttt{3,520} augmented samples in total. All training, testing, and
evaluation images were mutually disjoint to avoid data leakage.

\subsection{Weight Analysis}

The YOLOv4-tiny model was trained for \texttt{40{,}000} iterations on an Ubuntu
20.04 system. Training converged to an average loss of \texttt{0.0509}. We evaluated snapshot weight files saved at
\texttt{10{,}000} iteration intervals to identify the checkpoint with the best performance.

Table~\ref{tab:weightComparison} compares key metrics---precision, recall,
F1-score, IoU, and mean average precision (mAP)---across checkpoints. The
\texttt{30{,}000}-iteration checkpoint (`yolov4-custom\_30000.weights')
achieved the highest mAP (83.55\%) and the best overall balance between
precision and recall, and was therefore selected as the deployed detector for
ThermoCAPTCHA.

\begin{table*}[!htp]
\footnotesize
\caption{Comparison of YOLOv4-tiny checkpoints saved every 10{,}000 iterations.}
\label{tab:weightComparison}
\centering
\begin{tabular}{c|c|c|c|c|c|c|c|c|c|c|c|c}
\toprule
\textbf{Weights} & \textbf{Detections} & \textbf{Unique} & \textbf{Conf.} &
\textbf{TP} & \textbf{FP} & \textbf{FN} & \textbf{Prec.} & \textbf{Recall} &
\textbf{F1} & \textbf{IoU} & \textbf{Avg.\ IoU} & \textbf{mAP} \\
 & \textbf{Count} & \textbf{Count} & \textbf{Thresh.} & & & &
(\%) & (\%) & (\%) & (\%) & (\%) & (\%) \\
\midrule
10{,}000 & 1,286 & 745 & 0.25 & 722 & 257 & 23 & 74.0 & 97.0 & 84.0 & 50 &
59.14 & 80.89 \\
20{,}000 & 2,343 & 745 & 0.25 & 716 & 319 & 29 & 69.0 & 96.0 & 80.0 & 50 &
53.77 & 78.53 \\
30{,}000 & 2,756 & 745 & 0.25 & 715 & 212 & 30 & 77.0 & 96.0 & 86.0 & 50 &
61.51 & \textbf{83.55} \\
40{,}000 & 3,695 & 745 & 0.25 & 695 & 529 & 50 & 57.0 & 93.0 & 71.0 & 50 &
44.88 & 74.82 \\
\bottomrule
\end{tabular}
\end{table*}

%-------------------------------------------------------------------------------
\section{Detailed FGSM Evaluation Results}
\label{AppendixFGSM}
%-------------------------------------------------------------------------------

This section expands on the adversarial robustness results summarized in
Section~\ref{subsec:rq2}. We evaluated FGSM perturbations across seven
distortion budgets using 20 thermal samples (10 human, 10 non-human).  
For each $\epsilon$, we report (i) evasion (human $\rightarrow$ non-human),
(ii) induction (non-human $\rightarrow$ human), and (iii) overall
misclassification rate. To illustrate the qualitative nature of these
perturbations, Table~\ref{tab:fgsm-images} presents adversarial examples from a
representative thermal sample.

\begin{table*}[t]
\centering
\caption{Representative FGSM adversarial examples across different $\epsilon$ values.  
Each column shows the perturbed thermal image (top), perturbation pattern (middle), and the corresponding difference map (bottom). Confidence values refer to the human-detection score output by the model.}
\label{tab:fgsm-images}

\begin{tabular}{c c c c c c c c}
\toprule
\textbf{$\epsilon$} & \textbf{0.01} & \textbf{0.02} & \textbf{0.03} & \textbf{0.10} & \textbf{0.20} & \textbf{0.30} & \textbf{0.50} \\
\midrule

\makecell{\textbf{Adversarial} \\ \textbf{Image}} &
\makecell{\includegraphics[width=0.11\textwidth]{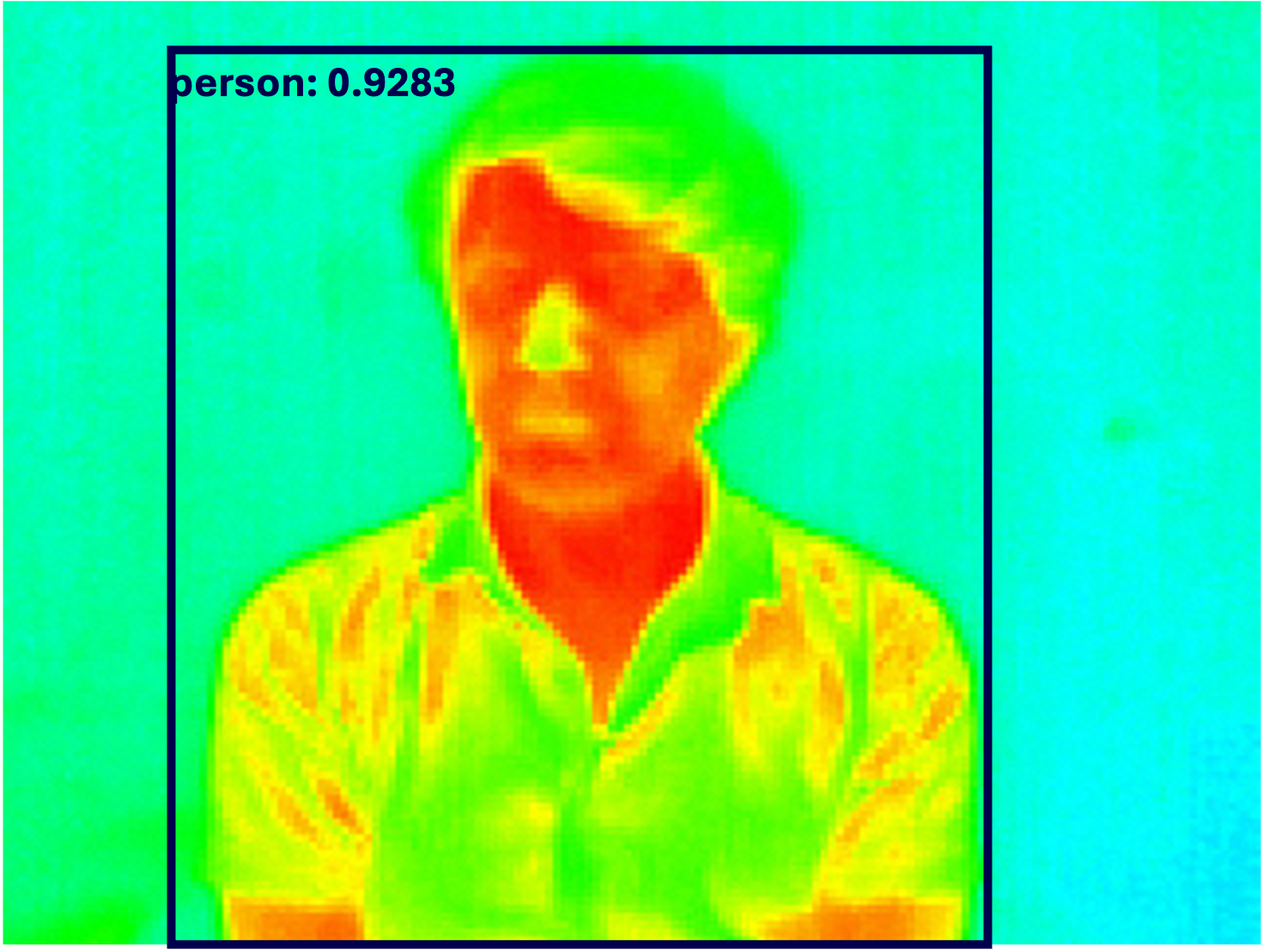} \\ \small\textcolor{darkgreen}{Conf: 0.9283}} &
\makecell{\includegraphics[width=0.11\textwidth]{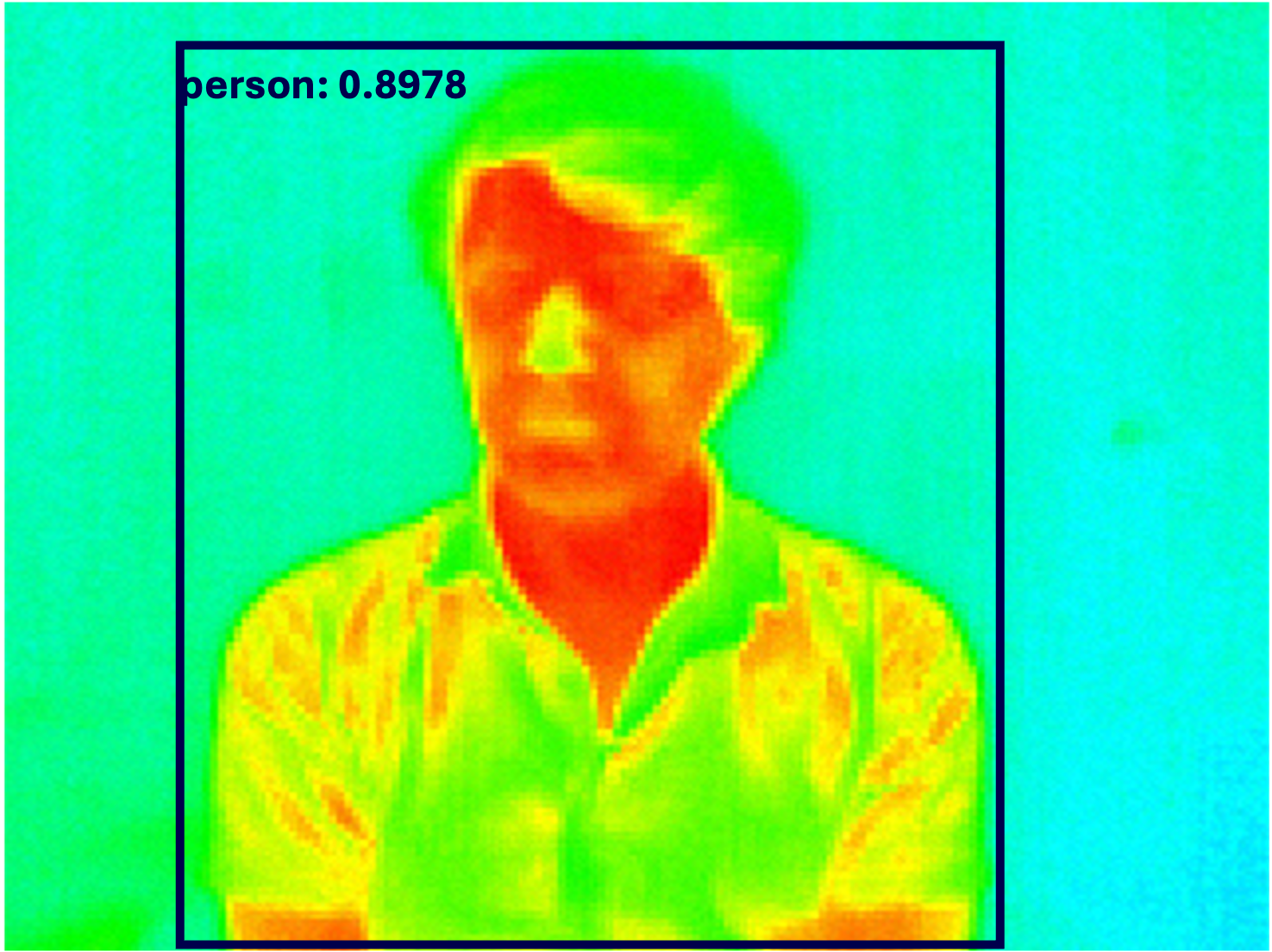} \\ \small\textcolor{darkgreen}{Conf: 0.8978}} &
\makecell{\includegraphics[width=0.11\textwidth]{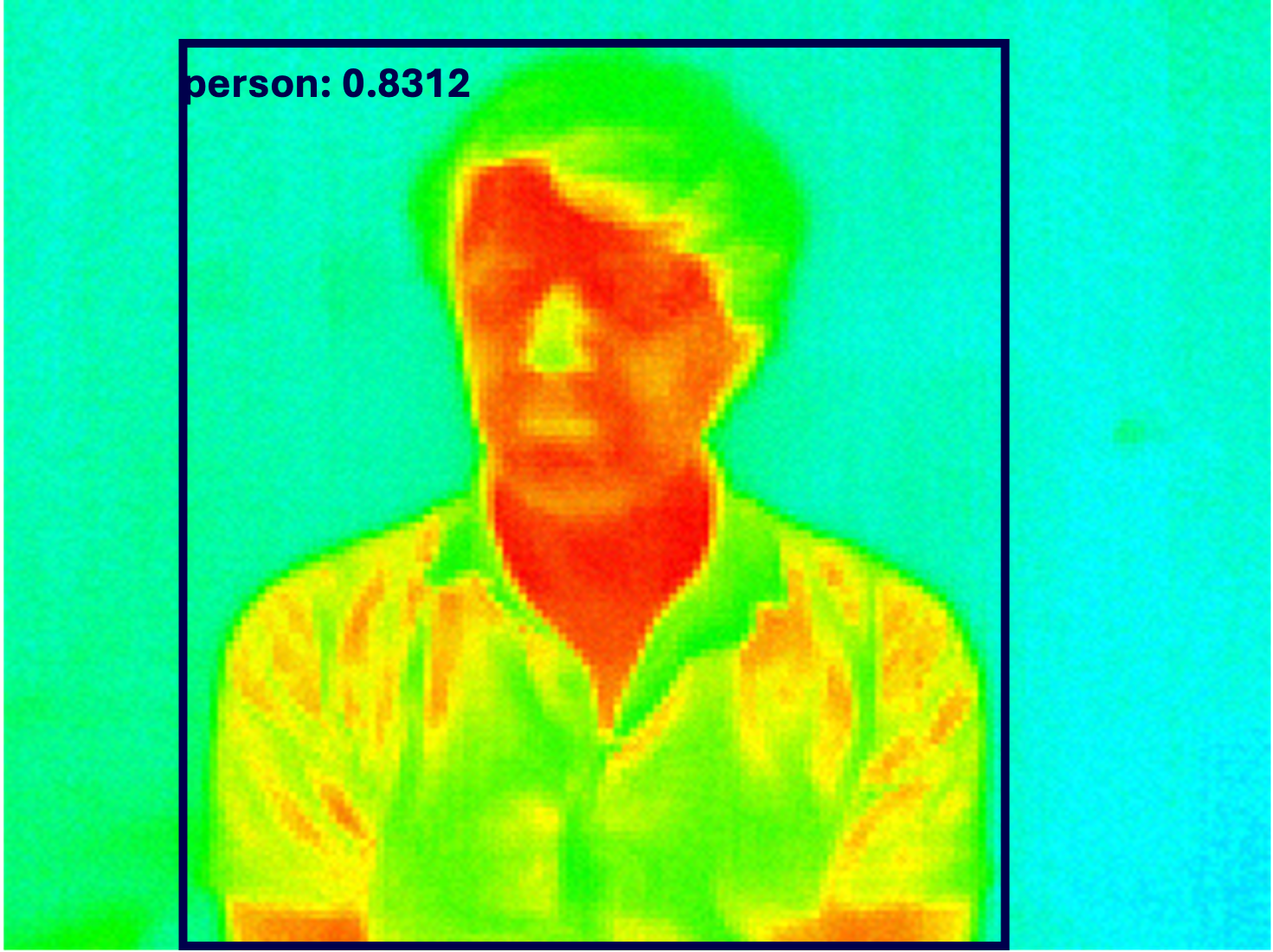} \\ \small\textcolor{darkgreen}{Conf: 0.8312}} &
\makecell{\includegraphics[width=0.11\textwidth]{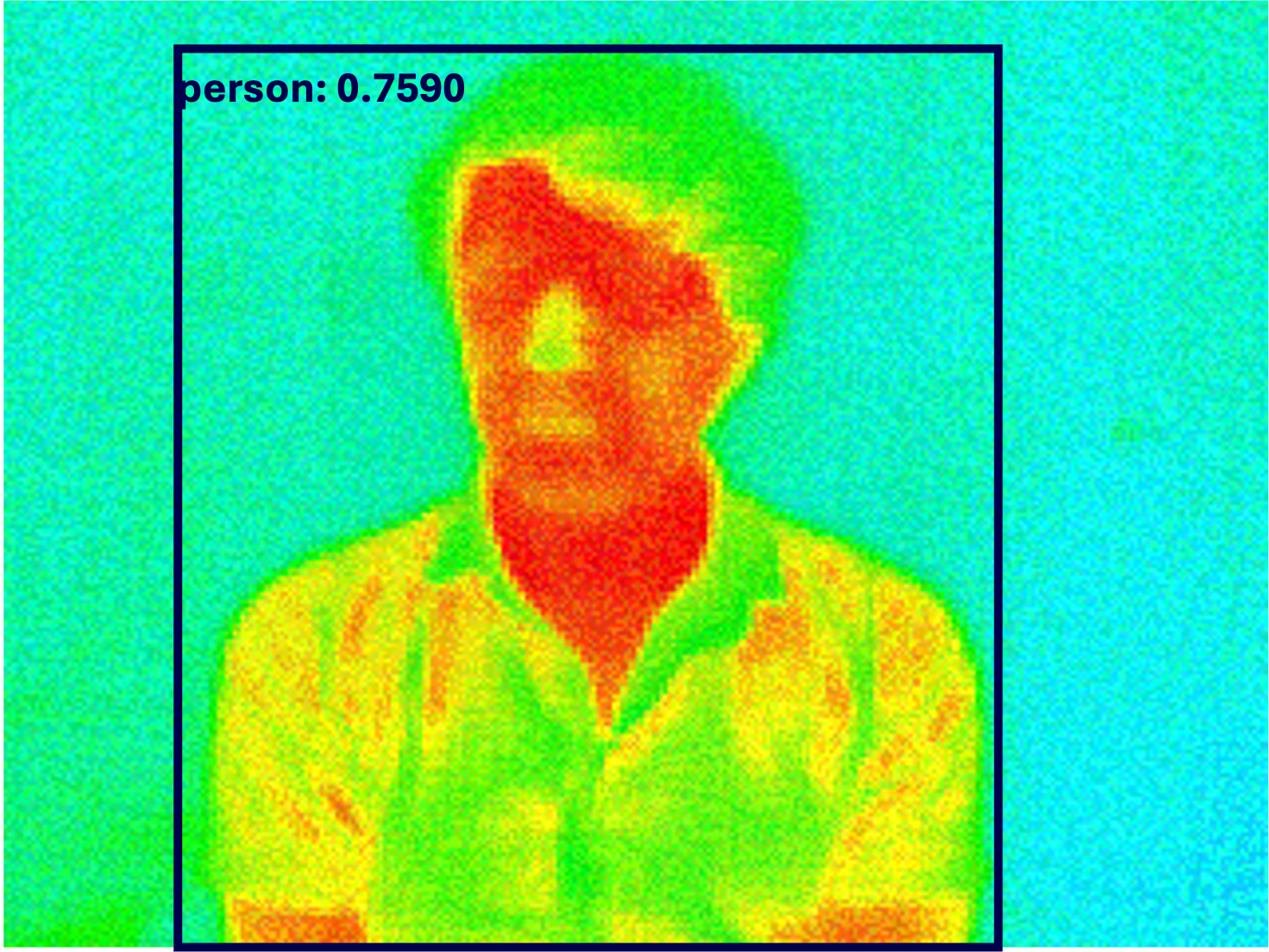} \\ \small\textcolor{darkgreen}{Conf: 0.7590}} &
\makecell{\includegraphics[width=0.11\textwidth]{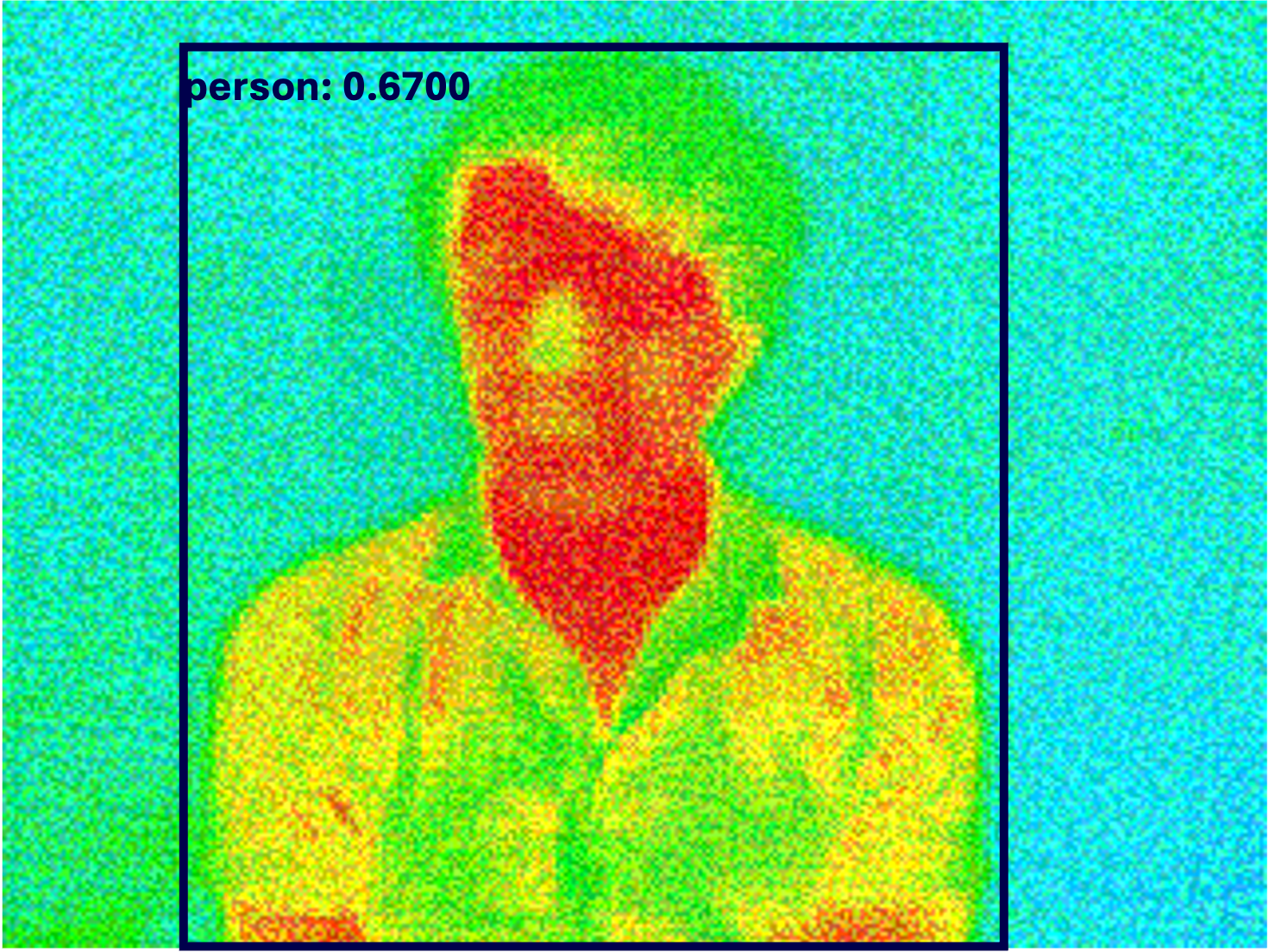} \\ \small\textcolor{darkgreen}{Conf: 0.6700}} &
\makecell{\includegraphics[width=0.11\textwidth]{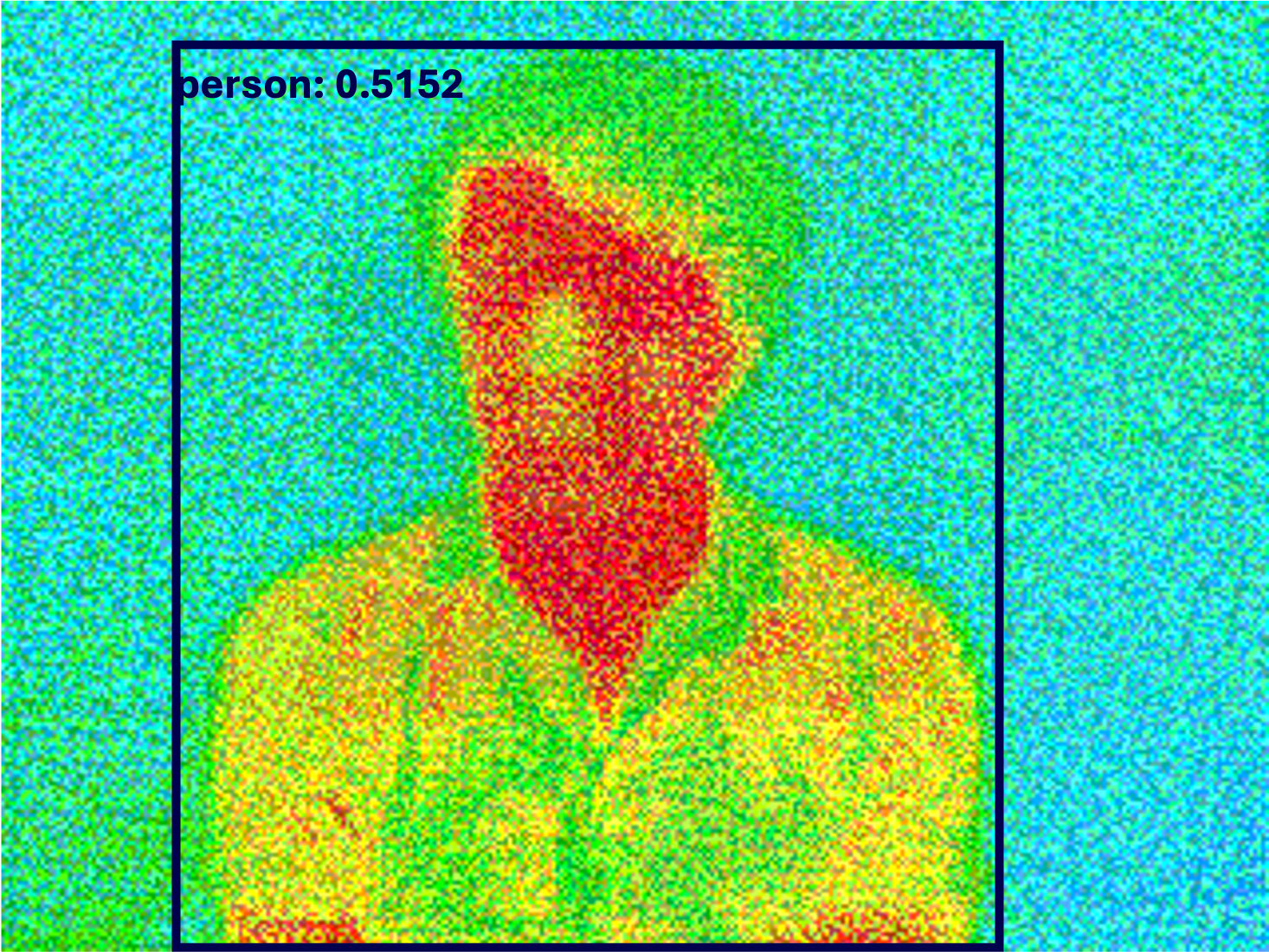} \\ \small\textcolor{darkgreen}{Conf: 0.5152}} &
\makecell{\includegraphics[width=0.11\textwidth]{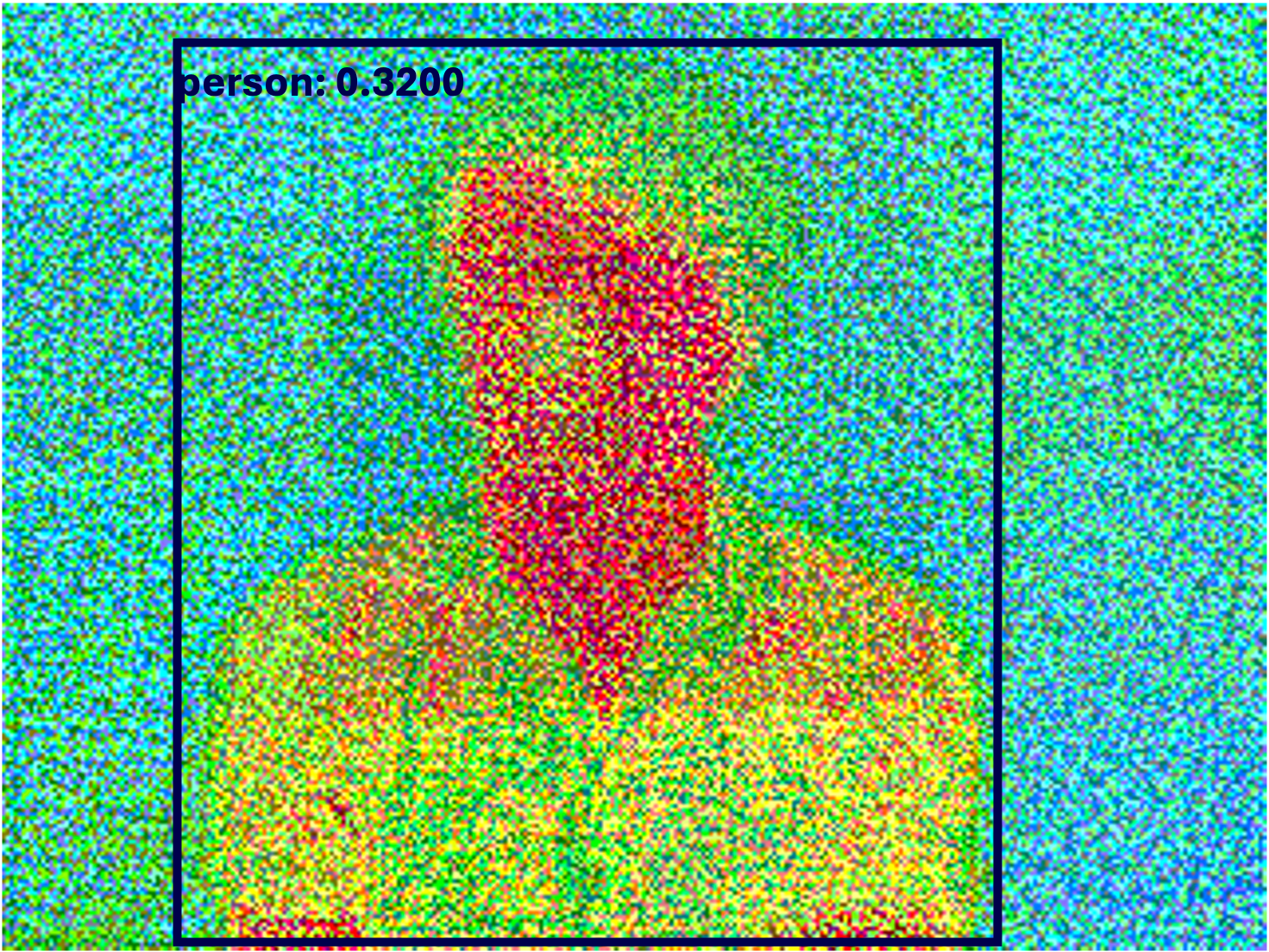} \\ \small\textcolor{darkred}{Conf: 0.3200}} \\
\midrule

\makecell{\textbf{Perturbation} \\ \textbf{Pattern}} &
\makecell{\includegraphics[width=0.11\textwidth]{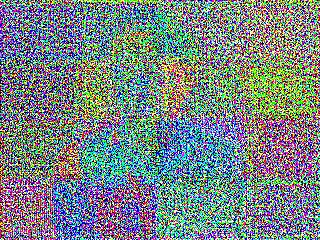}} &
\makecell{\includegraphics[width=0.11\textwidth]{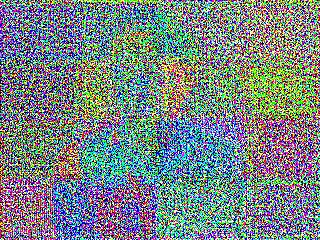}} &
\makecell{\includegraphics[width=0.11\textwidth]{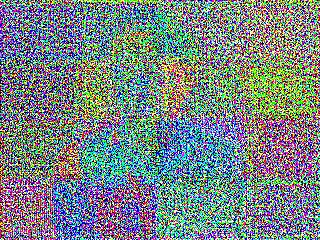}} &
\makecell{\includegraphics[width=0.11\textwidth]{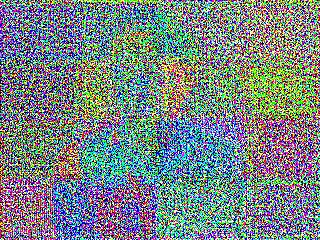}} &
\makecell{\includegraphics[width=0.11\textwidth]{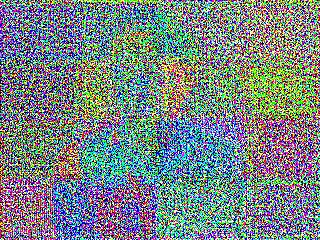}} &
\makecell{\includegraphics[width=0.11\textwidth]{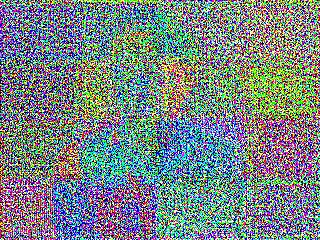}} &
\makecell{\includegraphics[width=0.11\textwidth]{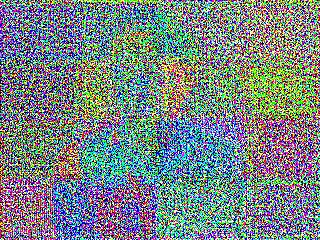}} \\
\midrule

\makecell{\textbf{Difference} \\ \textbf{Map}} &
\makecell{\includegraphics[width=0.11\textwidth]{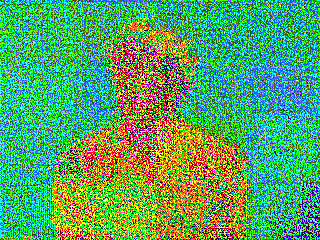}} &
\makecell{\includegraphics[width=0.11\textwidth]{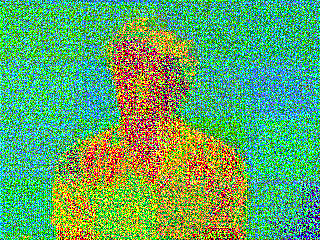}} &
\makecell{\includegraphics[width=0.11\textwidth]{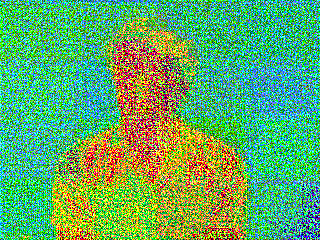}} &
\makecell{\includegraphics[width=0.11\textwidth]{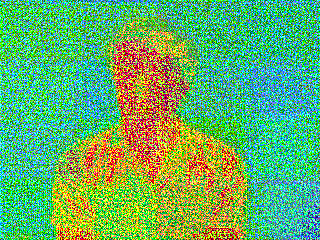}} &
\makecell{\includegraphics[width=0.11\textwidth]{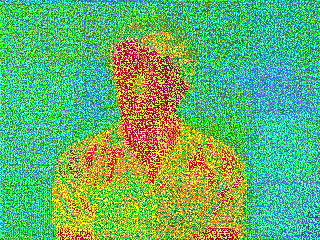}} &
\makecell{\includegraphics[width=0.11\textwidth]{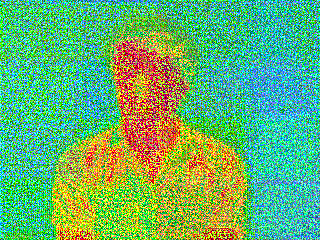}} &
\makecell{\includegraphics[width=0.11\textwidth]{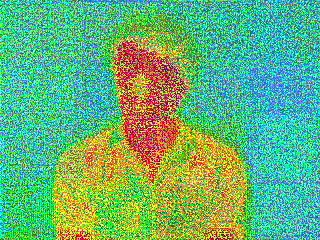}} \\

\bottomrule
\end{tabular}
\end{table*}

Table~\ref{tab:fgsm-detailed} shows the quantitative misclassification outcomes
across all 20 samples. Perturbations with $\epsilon \le 0.10$ produced no
misclassifications, while substantial degradation occurred only at large
distortion budgets ($\epsilon \ge 0.30$) that introduce visually unrealistic
artifacts.

\begin{table}[h!]
\centering
\caption{FGSM misclassification rates across 20 thermal samples.}
\label{tab:fgsm-detailed}
\resizebox{\columnwidth}{!}{
\begin{tabular}{l c c c c}
\toprule
\textbf{$\epsilon$} &
\textbf{Evasion} &
\textbf{Induction} &
\textbf{Success (\%)} &
\textbf{Miscls.} \\
\midrule
0.01 & 0\% & 0\% & 0\% & 0/20 \\
0.02 & 0\% & 0\% & 0\% & 0/20 \\
0.03 & 0\% & 0\% & 0\% & 0/20 \\
0.10 & 0\% & 0\% & 0\% & 0/20 \\
0.20 & 10\% & 0\% & 5\% & 1/20 \\
0.30 & 40\% & 20\% & 30\% & 6/20 \\
0.50 & 90\% & 70\% & 80\% & 16/20 \\
\bottomrule
\end{tabular}
}
\end{table}

%---------------------------------------------------------------
\section{Proof of Token Reuse Infeasibility}\label{Proof}
%----------------------------------------------------------------

We formally show that a traceable token issued by the ThermoCAPTCHA cannot
be reused by an attacker, even when a legitimate CAPTCHA-farm worker \(W\)
forwards the exact ciphertext to an adversary \(A\). The security argument
relies on the binding of the token to session- and device-specific values and
on the unforgeability of the MAC under the shared secret key.

\paragraph{Notation.}
Let:
\begin{itemize}
    \item \(T\): the traceable token,
    \item \(SK_{\text{server}}\): CAPTCHA server secret key (used for encryption),
    \item \(K_{\text{shared}}\): shared secret between the web server and CAPTCHA service,
    \item \(sess_{id}\): session identifier,
    \item \(dev_{fp}\): device fingerprint,
    \item \(nonce\): freshness nonce,
    \item \(exp\): expiration timestamp,
    \item \(UID\): user identifier (e.g., IP, public-key hash).
\end{itemize}

\paragraph{Token Construction.}
The server constructs a token of the form:
\begin{equation}
\label{eq:token-construct}
T = \text{Enc}_{SK_{\text{server}}}
\left(
    \text{MAC}_{K_{\text{shared}}}
    \big(
        UID \parallel sess_{id} \parallel dev_{fp} \parallel
        nonce \parallel exp
    \big)
\right).
\end{equation}

\paragraph{Attack Scenario.}
Assume a worker \(W\) legitimately solves a CAPTCHA and receives:
\begin{equation}
\label{eq:token-worker}
\begin{aligned}
T_W
= \text{Enc}_{SK_{\text{server}}}
\Big(
    \text{MAC}_{K_{\text{shared}}}
    \big(
        &UID_W \parallel sess_{id}^W \parallel dev_{fp}^W \\
        &\parallel~nonce^W \parallel exp^W
    \big)
\Big).
\end{aligned}
\end{equation}

The worker forwards \(T_W\) to an attacker \(A\), who issues a verification
request from a different device, session, and network context. The CAPTCHA
server decrypts \(T_W\) and recomputes the MAC on:
\begin{equation}
\label{eq:token-attack}
\begin{aligned}
\text{MAC}_{K_{\text{shared}}}
\big(
    &UID_A \parallel sess_{id}^A \parallel dev_{fp}^A \\
    &\parallel~nonce^W \parallel exp^W
\big).
\end{aligned}
\end{equation}

\paragraph{Mismatch of Context-Bound Values.}
Forwarding necessarily changes at least one of:
\[
(UID,~sess_{id},~dev_{fp}),
\]
so we have:
\[
(UID_A, sess_{id}^A, dev_{fp}^A)
\neq
(UID_W, sess_{id}^W, dev_{fp}^W).
\]

\paragraph{Failure of MAC Verification.}
From the unforgeability of MACs under \(K_{\text{shared}}\), it follows that:
\[
\text{MAC}_{K_{\text{shared}}}(A)
\ne
\text{MAC}_{K_{\text{shared}}}(W),
\]
so verification of \(T_W\) in attacker context \(A\) always fails.

\paragraph{Conclusion.}
The token \(T_W\) cannot be validated outside the original execution context.
Thus, the probability that attacker \(A\) successfully reuses a forwarded
token is:
\[
\Pr[\text{Reuse of token } T_W \text{ by attacker } A] = 0,
\]
under standard assumptions on the security of the encryption and MAC
primitives.

%-------------------------------------------------------------------------------
\section{User Study Methodology and Usability Statistics}
\label{usabilityMethodandStatistics}
%-------------------------------------------------------------------------------
This appendix provides additional methodological details and extended usability
statistics for the comparative study between \textsc{ThermoCAPTCHA} and Google
reCAPTCHA described in Section~\ref{userStudyMethodology}. The main text
summarizes the participant population, core procedure, and key quantitative
results; here we expand on (i) the study protocol and ethical safeguards, and
(ii) detailed subjective usability outcomes.

\subsection{Additional Methodological Details}

As noted in Section~\ref{userStudyMethodology}, each participant solved three
\textsc{ThermoCAPT-CHA} challenges and three reCAPTCHA challenges. Both systems
were deployed on the lab website and accessed via Google Forms. The order of
\textsc{ThermoCAPTCHA} and reCAPTCHA challenges was randomized to reduce
ordering effects.

To maintain ethical standards, each participant was presented with three
documents before participation:
\begin{itemize}
    \item a CITI Program certificate verifying that the research team completed
          Human Subjects Research training,
    \item a flyer explaining the study goals and overall procedure, and
    \item an informed-consent form describing risks, data handling practices,
          and the right to withdraw at any time.
\end{itemize}

The study took place in a controlled lab environment and was overseen by an
undergraduate surveyor who was not involved in the design or implementation of
\textsc{ThermoCAPTCHA}, in order to minimize experimenter bias. Participants
completed the tasks using Google Chrome in incognito mode, and the Google Form
did not require login, preserving anonymity.

When solving \textsc{ThermoCAPTCHA}, participants were prompted by the browser
to grant camera access. Because some browsers cache camera permissions across
page loads, cookies and site data were cleared between sessions to ensure a
consistent permission workflow for all users. Both interfaces followed Web
Content Accessibility Guidelines (WCAG)~\cite{WebConte28:online}.

The questionnaire layout (referenced in Section~\ref{userStudyMethodology})
consisted of:
\begin{itemize}
    \item an initial one-page guide explaining how to solve each challenge in
          \textsc{ThermoCAPTCHA} and reCAPTCHA,
    \item three distinct \textsc{ThermoCAPTCHA} challenges,
    \item three distinct reCAPTCHA challenges,
    \item a short questionnaire on familiarity with CAPTCHAs and experience
          with face-authentication systems (e.g., blinking, smiling,
          liveness checks), and
    \item a final questionnaire asking users to rate \textsc{ThermoCAPTCHA}
          on comfort, fun, and ease of use relative to reCAPTCHA.
\end{itemize}

For subjective evaluation, we used Likert-style questions rather than the
System Usability Scale (SUS), following W3C accessibility recommendations and
prior work on blind and visually challenged users~\cite{Inaccess90:online,fanelle2020blind}.
Participants rated \textsc{ThermoCAPTCHA} and reCAPTCHA on fun and ease of use
using a 5-point scale:
\begin{itemize}
    \item 1--2: reCAPTCHA is more fun/easy,
    \item 3: both systems are equally fun/easy,
    \item 4--5: \textsc{ThermoCAPTCHA} is more fun/easy.
\end{itemize}
Separate 1--5 scales were used for overall comfort and for willingness to adopt
\textsc{ThermoCAPTCHA} in real deployments.

\subsection{Extended Usability Statistics}

Table~\ref{tab:averageTimeAndAccuracy} in the main text reports the Correct
Attempt Ratio (CAR) and completion times across user groups and conditions.
Here we provide additional subjective statistics that complement those
quantitative results.

\begin{figure*}[t]
\centering
\Description{Timing distribution of each system for both normal and visually challenged users.}
\subfigure[normal users]{
\includegraphics[width=0.50\textwidth]{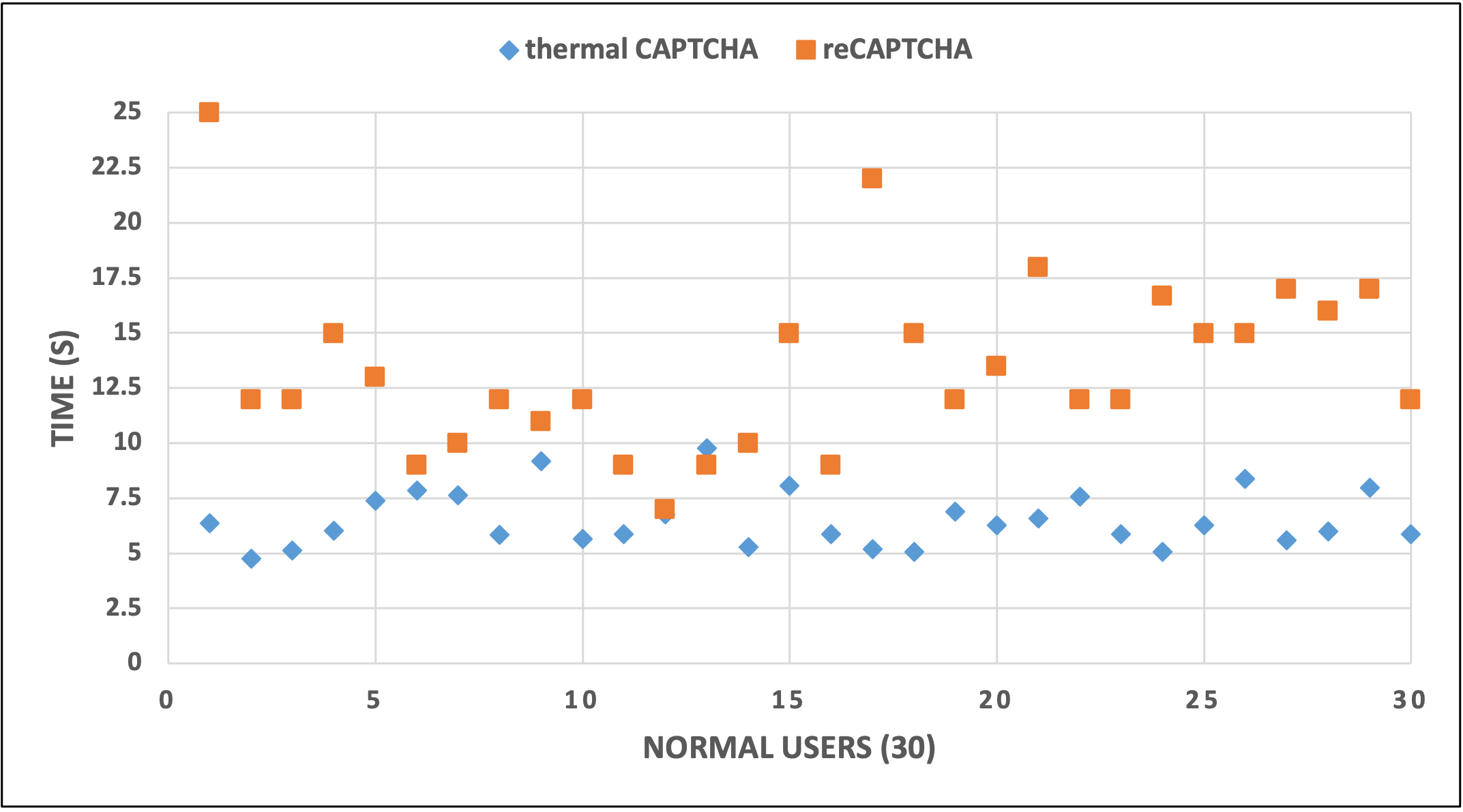}
}
\subfigure[visually challenged users]{
\includegraphics[width=.40\textwidth]{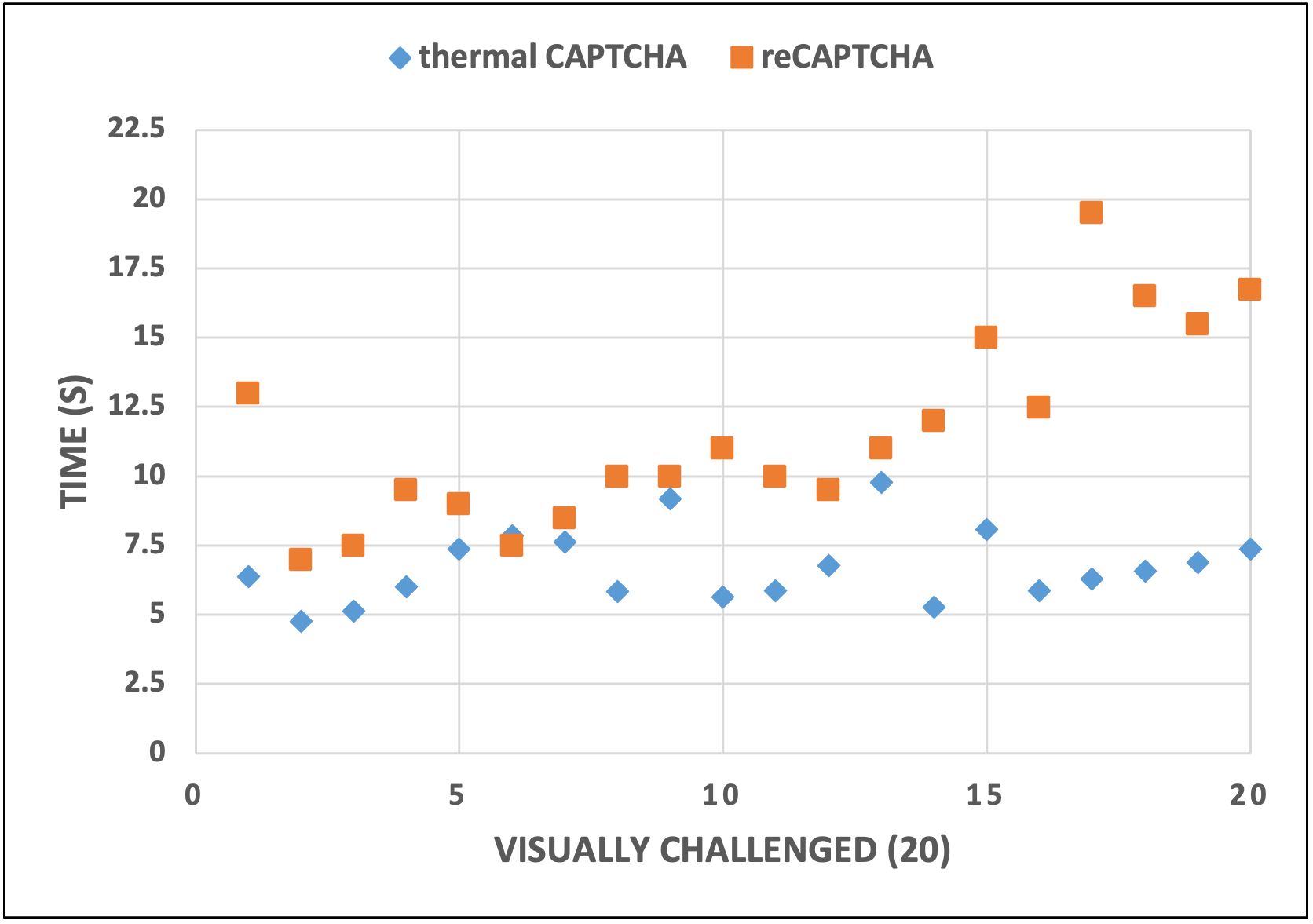}
}
\caption{Timing distribution of \textsc{ThermoCAPTCHA} and reCAPTCHA for sighted and visually challenged users.} 
\label{fig:completionTime}
\end{figure*}

\begin{figure}[htp!]
\centering
\Description{Comparison between reCAPTCHA and \textsc{ThermoCAPTCHA} on fun and ease-of-use ratings for sighted and visually challenged users.}
\subfigure[normal users]{
    \includegraphics[width=0.47\textwidth]{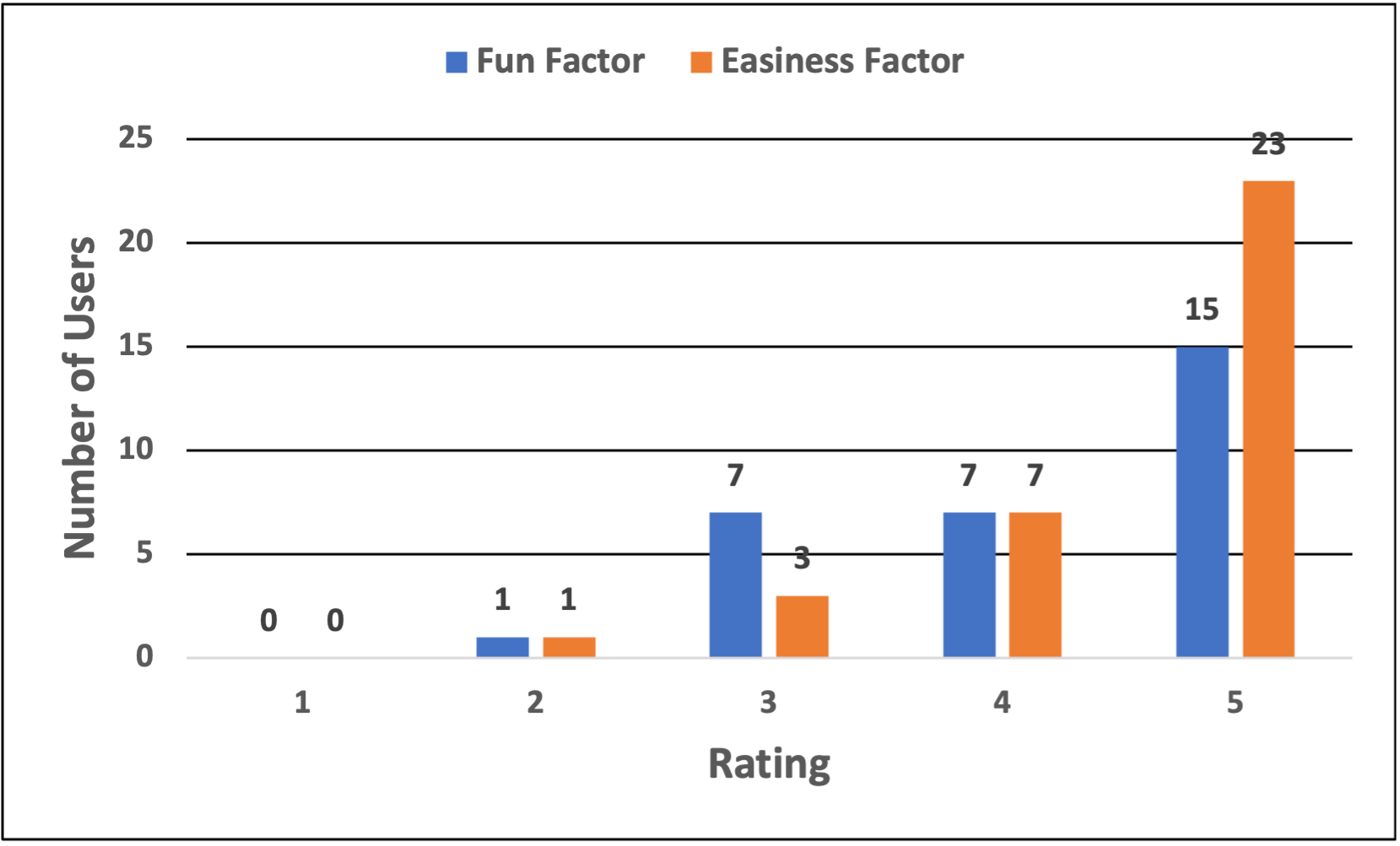}
    \label{fig:funAndEasiness1}
}
\subfigure[visually challenged users]{
    \includegraphics[width=0.47\textwidth]{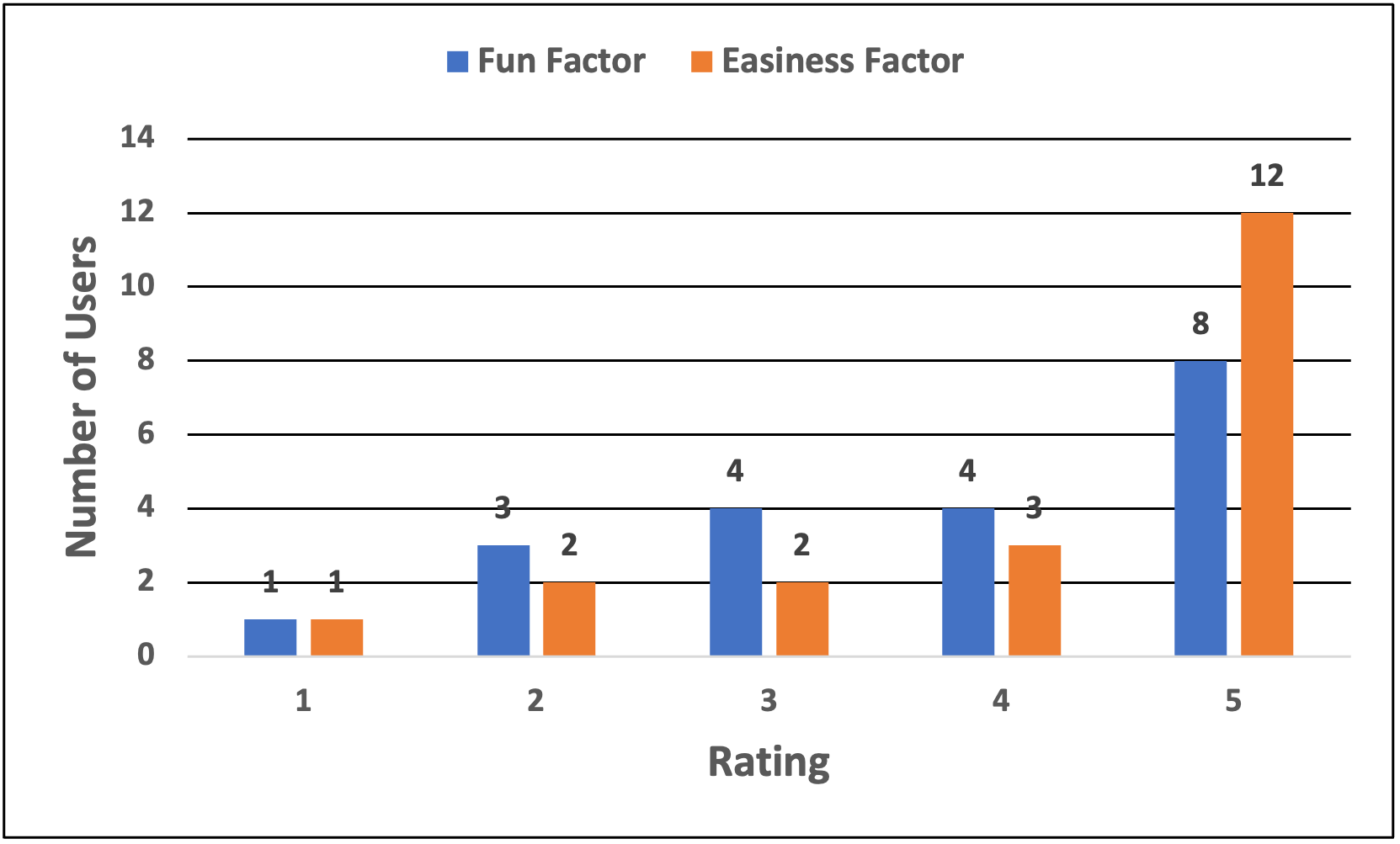}
    \label{fig:funAndEasiness2}
}
\caption{Fun and ease-of-use ratings for \textsc{ThermoCAPTCHA} vs.\ reCAPTCHA.
Ratings 1--2 indicate a preference for reCAPTCHA, 3 indicates no preference,
and 4--5 indicate a preference for \textsc{ThermoCAPTCHA}.}
\label{fig:funAndEasiness}
\end{figure}

After completing all six challenges, participants rated fun and ease of use for
both systems. Figure~\ref{fig:funAndEasiness} breaks down these responses by
user group.

For normal users, 73.30\% of responses fell in the 4--5 range, indicating that
participants generally found \textsc{ThermoCAPTCHA} at least somewhat more fun
and easier to use than reCAPTCHA. Among visually challenged users, 60\% of
responses were in the 4--5 range, with relatively few ratings in the 1--2
(reCAPTCHA-preferred) region. These distributions align with the summary
statistics in Section~\ref{userStudyMethodology}, where 86.67\% of sighted
users and 75\% of visually challenged users rated \textsc{ThermoCAPTCHA} as
easier than or superior to reCAPTCHA.

\begin{figure}[ht]
\centering
\Description{User comfort level statistics and reasons for preferring \textsc{ThermoCAPTCHA}.}
\subfigure[Comfort levels]{
    \includegraphics[width=0.47\textwidth]{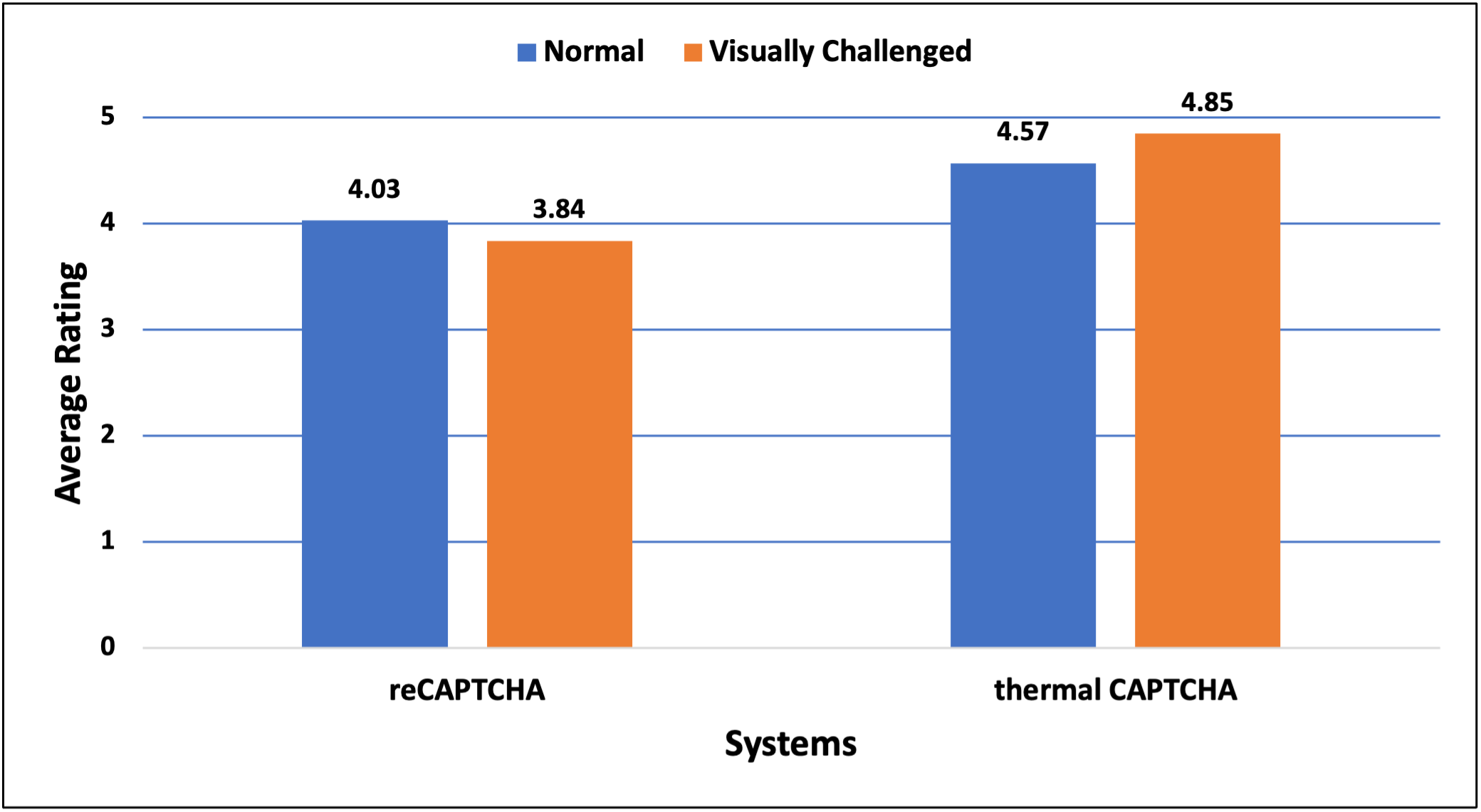}
    \label{fig:comfortLevel}
}
\subfigure[Self-reported reasons]{
    \includegraphics[width=0.47\textwidth]{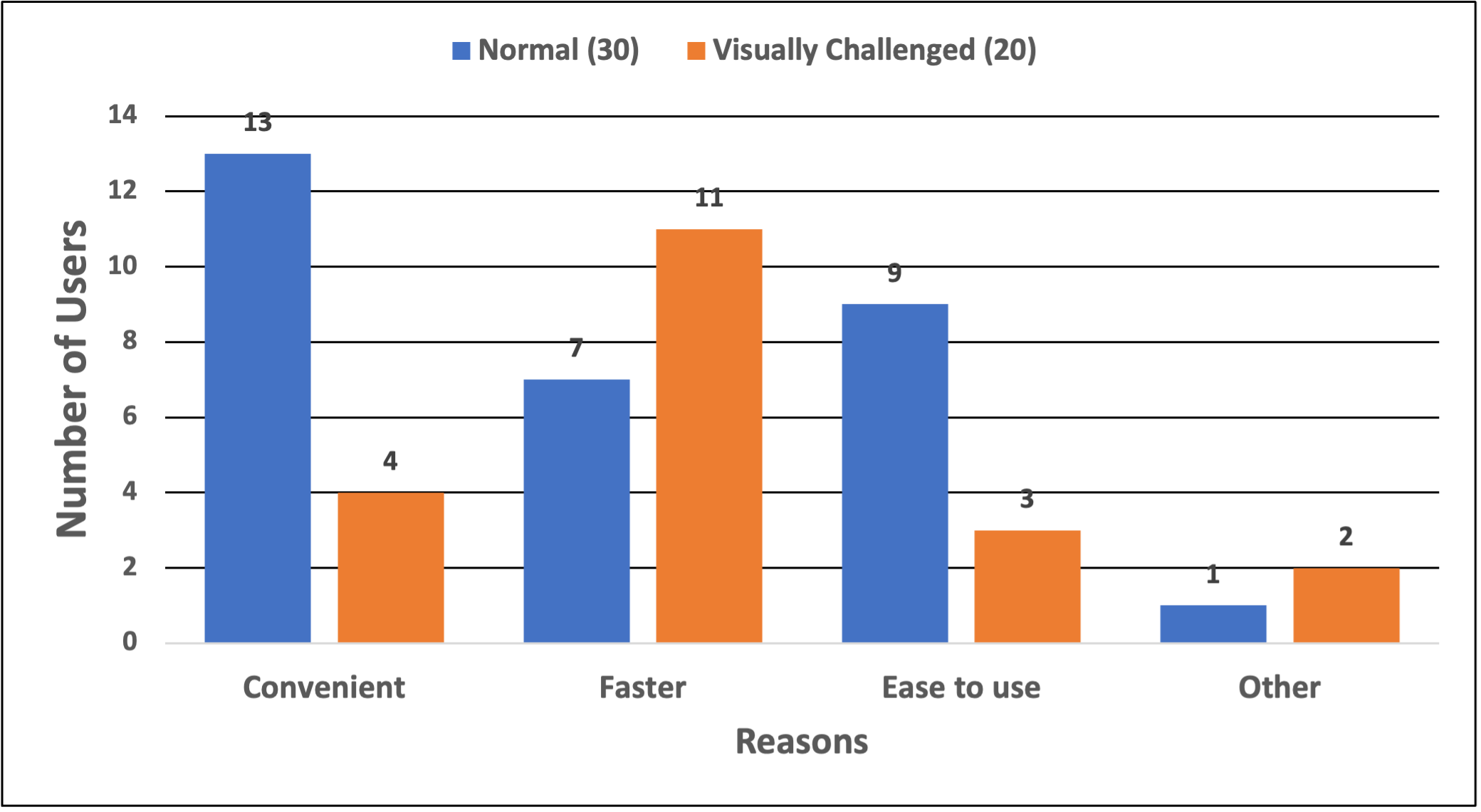}
    \label{fig:comfortReason}
}
\caption{User comfort levels and self-reported reasons for preferring \textsc{ThermoCAPTCHA} over reCAPTCHA.}
\label{fig:comfortLevelandReson}
\end{figure}

We also collected detailed comfort ratings and self-reported reasons for
preferring (or not preferring) \textsc{ThermoCAPTCHA}. Figure~\ref{fig:comfortLevelandReson}
shows the comfort distribution and the primary reasons cited.

Both sighted and visually challenged users reported higher comfort with
\textsc{ThermoCAPTCHA} than with reCAPTCHA. Among normal users, 43\% selected
convenience as their primary reason for preferring \textsc{ThermoCAPTCHA}, and
30\% highlighted ease of use. Among visually challenged users, 55\% cited speed
as the dominant factor, with additional responses emphasizing reduced visual
demand and lower frustration. These subjective results are consistent with the
objective CAR and timing improvements reported in the main text, and together
they support the conclusion that \textsc{ThermoCAPTCHA} offers both improved
task performance and favorable user experience, particularly for users with
visual impairments.

\end{document}